\documentclass[draft]{agujournal2019}
\usepackage{url} %this package should fix any errors with URLs in refs.
\usepackage{lineno}
\usepackage[inline]{trackchanges} %for better track changes. finalnew option will compile document with changes incorporated.
\usepackage{soul}
\usepackage{multirow}
\usepackage{amsmath,amssymb}
\draftfalse

\journalname{JGR-Space Physics}

\begin{document}

\title{The Structure and Kinetic Ion Behavior of Low Mach Number Shocks}

\authors{D. B. Graham \affil{1}, Yu. V. Khotyaintsev\affil{1}}

\affiliation{1}{Swedish Institute of Space Physics, Uppsala, Sweden.}

\correspondingauthor{D. B. Graham}{dgraham@irfu.se}

\begin{keypoints}
\item For comparable low Mach number shocks, the width $l$ increases with decreasing shock-normal angle $\theta_{Bn}$. 
\item Increasing $l$ suppresses proton reflection at the ramp; decreasing $\theta_{Bn}$ leads to more rapid gyrophase mixing of reflected protons.
\item Gyrophase bunching of alpha particles leads to compressional magnetic field fluctuations downstream of the shocks. 
\end{keypoints}

%\date{\today}
\begin{abstract}
Low Mach number collisionless shocks are routinely observed in the solar wind and upstream of planetary bodies. However, most in situ observations have lacked the necessary temporal resolution to directly study the kinetic behavior of ions across these shocks. We investigate a series of five low Mach number bow shock crossings observed by the Magnetospheric Multiscale (MMS) mission. The five shocks had comparable Mach numbers, but varying shock-normal angles ($66^{\circ} \lesssim \theta_{Bn} \lesssim 89^{\circ}$) and ramp widths ($5~\mathrm{km} \lesssim l \lesssim 100~\mathrm{km}$). The shock width is shown to be crucial in determining the fraction of protons reflected and energized by the shock, with proton reflection increasing with decreasing shock width. As the shock width increases proton reflection is arrested entirely. For nearly perpendicular shocks, reflected protons exhibit quasi-periodic structures, which persist far downstream of the shock. As the shock-normal angle becomes more oblique these periodic proton structures broaden to form an energetic halo population. Periodic fluctuations in the magnetic field downstream of the shocks are generated by fluctuations in dynamic pressure of alpha particles, which are decelerated by the cross-shock potential and subsequently undergo gyrophase bunching. These results demonstrate that complex kinetic-scale ion dynamics occur in low Mach number shocks, which depend significantly on the shock profile. 
\end{abstract}

\section*{Plain Language Summary}
One of the most significant challenges when investigating shock waves in plasma with spacecraft data is having sufficiently high temporal resolution fields and particle measurements to resolve all the features of the shock. Typically, measurements of electric and magnetic fields are fast so the profiles of these fields can be studied in detail. In contrast, particle measurements are much slower, so the behavior of ions across shocks is more difficult to study. We use the Magnetospheric Multiscale (MMS) spacecraft, which provide high temporal resolution ion measurements to study the behavior of ions five across low Mach number shocks, where the magnetic field profiles of the shocks are relatively simple. The shocks varied in width and angle between the upstream magnetic field and the direction normal to the shock, the shock-normal angle. We show that despite the simple magnetic field profile of these shocks, complex ion dynamics occur, such as proton reflection off the shock boundary. The ion dynamics depend strongly on the shock width and shock-normal angle. These results show that the behavior of ions across the shock depend strongly on the shock profile, and that while some shocks may appear simple complex particle behavior is likely occurring. 

%\begin{article}

\section{Introduction}
Collisionless shocks are a fundamental process in plasma physics throughout the universe, and are responsible for intense particle energization. Collisionless shocks redistribute upstream flow energy to heat ions and electrons across the shock. In our solar system, collisionless shocks are typically observed in the solar wind, such as associated with Coronal Mass Ejections (CMEs) \cite{webb2012} and co-rotation interaction regions \cite{smith1976}, and as bow shocks of planetary bodies \cite{Balogh2013}. Of these shocks, Earth's bow shock is the most studied using spacecraft observations. At Earth's bow shock the super-Alfv\'{e}nic solar wind is compressed and heated to form Earth's magnetosheath, which convects around the magnetosphere. 

Typically, Earth's bow shock is characterized by supercritical shocks \cite{lalti2022a}, with moderate Mach numbers exceeding the critical Mach number \cite{kennel1985}. These shocks cannot be sustained by resistivity alone, such as by wave-particle interactions \cite{wilson2014,wu1984}. As a result these shocks reflect a significant portion of incoming solar wind ions. For quasi-perpendicular shocks, these reflected ions are accelerated by the solar wind convection electric field and cyclotron turned by the magnetic field, and subsequently cross the shock. These shocks tend to be complex and dynamic \cite{johlander2023}.  

In contrast, low Mach number subcritical shocks tend to be much more laminar \cite{greenstadt1975,formisano1971}. Subcritical shocks are often observed in the solar wind \cite{russell2009}. For these shocks resistivity is sufficient to sustain the shocks, without ion reflection, although a small amount of proton reflection at some low Mach number shocks has been reported \cite{sckopke1983,greenstadt1987}. Observations have shown that the magnetic field profiles of these shocks are quite laminar \cite{greenstadt1975,formisano1971}, although upstream and downstream fluctuations are frequently observed. Observations have shown that precursor whistler waves are routinely observed upstream of low Mach number shocks \cite{fairfield1975,farris1993}. These whistlers are often phase-standing with the shock ramp. However, observations have also shown that the whistler waves can have properties inconsistent with phase-standing waves \cite{wilson2017}. 

Downstream of low Mach number shocks, compressional fluctuations in the magnetic field have been observed \cite{farris1993,russell2009}. These fluctuations were investigated by \citeA{balikhin2008} using Venus Express, who argued that gyrophase bunching of protons downstream of the shock was responsible. Using THEMIS, \citeA{pope2019} found that the downstream fluctuations were consistent with gyrophase bunching of protons and alpha (He$^{2+}$) particles. The fluctuations in pressure associated with gyrophase bunching were argued produce fluctuations in the ion pressure, which decay away from the shock as the pressure fluctuations decrease \cite{zilbersher1998, gedalin2015}. The compression magnetic field fluctuations occur to maintain pressure balance downstream of the shocks. 

The solar wind is primarily composed of electrons, protons, and He$^{2+}$ particles. Nominally, the ratio of He$^{2+}$ to proton number densities are $\approx 0.04$, although this number varies with solar activity \cite{ogilvie1969,elliott2018}. Across the bow shock protons and He$^{2+}$ are expected to undergo different motions \cite{gedalin2017}, due to the distinct mass-to-charge rations. Observations have shown that downstream of the bow shock, He$^{2+}$ often form ring- and shell-like distributions \cite{fuselier1994}. Modelling and simulations suggest that He$^{2+}$ can modify the structure of collisionless shocks \cite{burgess1989,gedalin2017,gedalin2018,preisser2020,ofman2021}.

In rare cases Earth's bow shock is subcritical when the solar wind Alfv\'{e}n speed is usually large, for instance, in magnetic clouds following Coronal Mass Ejections. Observations of the low Mach number bow shock have shown that a small fraction of protons can be reflected by these shocks. Recently, \citeA{graham2024} investigated a nearly-perpendicular low Mach number bow shock. They found that proton reflection occurred for this shock and resulted in complex quasi-periodic proton structures persisting downstream of the shock. In contrast, upstream He$^{2+}$ and He$^{+}$ were directly transmitted across the shock ramp without reflection, but underwent periodic fluctuations downstream of the shock. 

In this paper, we expand on the work of \citeA{graham2024} to investigate the five bow shock crossings observed by the Magnetospheric Multiscale (MMS) spacecraft \cite{burch2016a} on 24 April 2023, while a magnetic cloud succeeding a Coronal Mass Ejection crossed Earth. The bow shock crossings were characterized by low Mach numbers and low plasma betas. As the magnetic cloud crossed Earth the shock-normal angle observed by MMS decreased with time from approximately perpendicular to oblique quasi-perpendicular shock geometries. The observations by the high time-resolution fields and particle instruments of MMS provide a unique opportunity to investigate the structure and kinetic ion behavior across low Mach number shocks as shock-normal angle changes. 

The paper is organized as follows: In section \ref{obanover}, we present the data used and provide an overview of the bow shock crossings and calculate their characteristics. In section \ref{sec4mod}, we investigate why the ion dynamics vary between the observed shocks, and with the aid of numerical modelling, we determine how the ion dynamics are governed by the shock width and shock-normal angle. We also investigate the low-frequency magnetic field perturbations observed downstream of these shocks. 
In section \ref{conclusionsec} we state the conclusions of this paper. 

\section{Observations and Overview} \label{obanover}
\subsection{MMS data}
We use data from the MMS spacecraft \cite{burch2016a}. Magnetic field ${\bf B}$ data are from Fluxgate Magnetometer (FGM) \cite{russell2016}. Electric field ${\bf E}$ are from Electric field Double Probes (EDP), which consists of the Spin-plane Double Probes (SDP) \cite{lindqvist2016} and Axial Double Probe (ADP) \cite{ergun2016}. The particle distributions and moments are measured using the Fast Plasma Investigation \cite{pollock2016}. 

We use fast survey and burst mode data. From FGM we use fast survey and burst mode ${\bf B}$ measurements sampled at 16 and 128~Hz, respectively. From EDP we use continuous burst mode ${\bf E}$ sampled at 8192~Hz. From FPI the ion and electron distributions and moments are sampled every 4.5~s in fast survey mode. Burst mode ion distributions and moments are sampled every 150~ms, and electron distributions and moments are sampled every 30~ms. The electron and ion distributions are measured in all directions with angular resolution of $11.25^{\circ}$. The energy resolution is $\Delta E/E \approx 0.13$ when the shocks were observed. 

For ion distributions this angular and energy resolution is too coarse to well resolve the cold ion beam of the solar wind, which makes the proton densities and temperatures very uncertain. Additionally, ions are sorted by energy per charge and FPI does not resolve different ion species. As a result solar wind He$^{2+}$ will be measured by FPI at higher energies (twice the proton energy when the particle speeds are equal). The proton moments are computed assuming the ion distributions consists of protons alone. As a result significant concentrations of He$^{2+}$ can modify the proton moments. In particular, in the solar wind significant concentrations of He$^{2+}$ will increase the proton density, bulk velocity and temperature moments when integrating over the entire ion distribution. In the solar wind the electron number density can be underestimated due to the core electron population not being fully resolved at low energies. Throughout this paper we use data from MMS1, unless otherwise stated. 

\subsection{Overview of the Shocks} \label{shockoverviewsec}
\begin{figure*}[htbp!]
\begin{center}
\includegraphics[width=160mm, height=130mm]{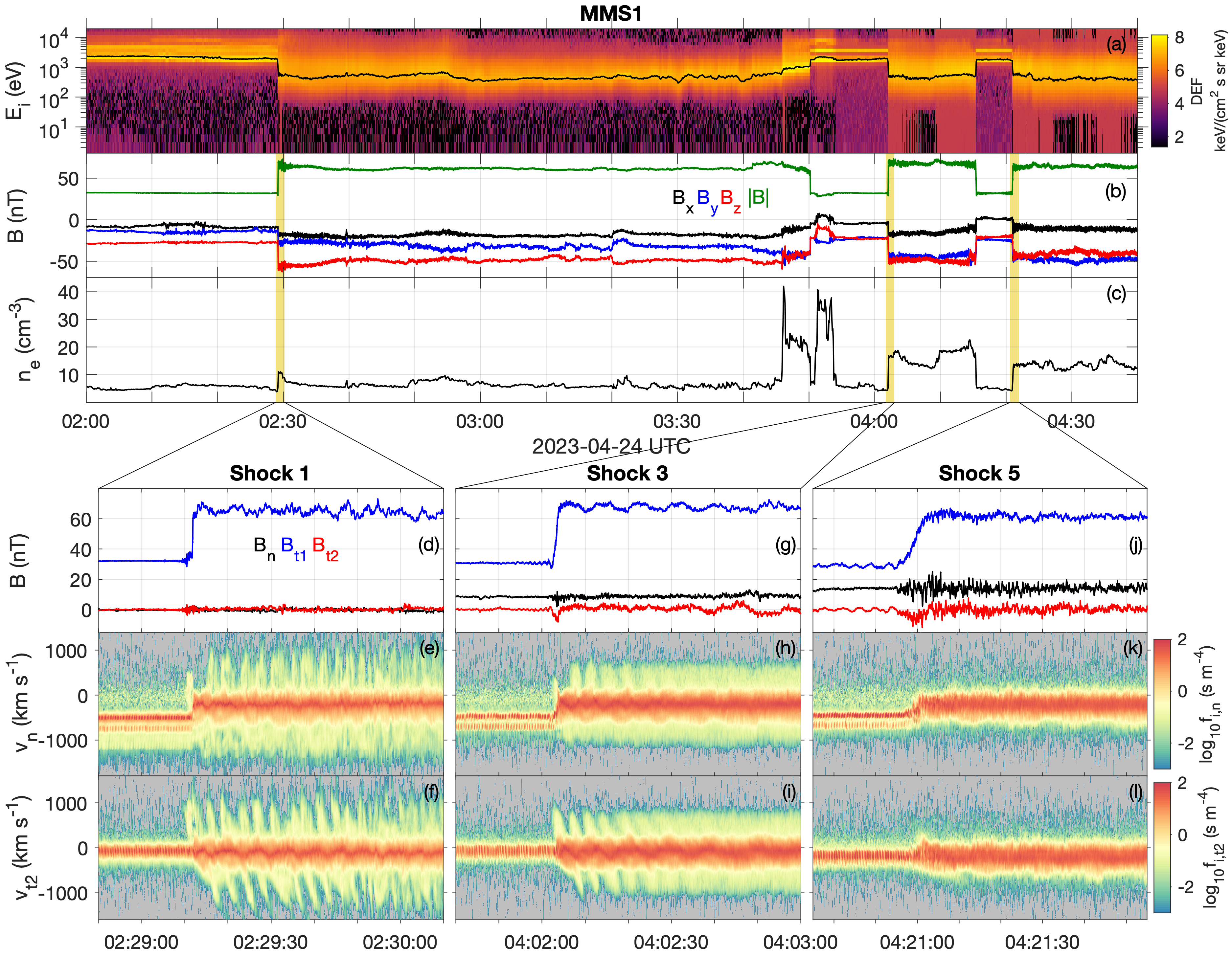}
\caption{Overview of the five bow shock crossings observed by MMS1 on 24 April 2023. 
Panels (a)--(c) show an extended interval including all the bow shock crossings. (a) Ion omnidirectional energy flux. (b) ${\bf B}$ in GSE coordinates. (c) Electron number density $n_e$. Panels (d)--(f), (g)--(i), and (j)--(l) show ${\bf B}$ and 1D reduced ion distributions across shocks 1, 3, and 5. 
(d), (g), and (j) ${\bf B}$ in local ($\hat{\bf n}$,$\hat{\bf t}_1$,$\hat{\bf t}_2$) shock coordinates. (e), (h), and (k) 1D reduced ion distributions along $v_n$. (f), (i), and (l) 1D reduced ion distributions along $v_{t2}$. The 1D reduced ion distributions are presented in the spacecraft frame. }
\label{overviewfig}
\end{center}
\end{figure*}

We now provide an overview of the five bow shock crossings observed by MMS on 24 April 2023. On 23 April 2023 at 17:34:32 UT a CME shock crossed MMS, while MMS was operating in slow survey mode in the solar wind. Following this, MMS observed the CME sheath then the magnetic cloud at $\sim$01:46:20 UT on 24 April 2023. Figures \ref{overviewfig}a--\ref{overviewfig}c show the ion omnidirectional energy flux, magnetic field ${\bf B}$ in Geocentric Solar Ecliptic (GSE) coordinates, and electron number density $n_e$. At the beginning of the interval MMS was located at (13.0,-17.0,-9.5)~$R_E$ and at (13.6,-15.3,-9.2)~$R_E$ at the end of the interval in GSE coordinates, where $R_E$ is Earth's radius. Between 02:00 and 04:30 UT the spacecraft crossed between the magnetic cloud behind the CME and the magnetosheath. The magnetic cloud plasma was characterized by cold ions and relatively constant ${\bf B}$, with $|{\bf B}| = 30$~nT. This corresponds to an unusually large $|{\bf B}|$ for the solar wind at $1$~AU, resulting in fast Alfv\'{e}n and magnetosonic speeds and a low Mach number bow shock. Similarly, ${\bf B}$ remains relatively constant in the magnetosheath intervals with $|{\bf B}| \approx 60$~nT, due to the low Mach number of the bow shock over this time interval. In the magnetosheath intervals the ion energy flux broadens in energy and peaks at lower energies due to the decrease in speed and increase in temperature as the ions cross the bow shock. For each of the bow shock crossings, $n_e$ increases by about a factor of $2$. 

Between 03:46 UT and 03:54 UT we observe a large enhancement in $n_e$. This corresponds to a pressure pulse inside the magnetic cloud, which is observed in conjunction with the second bow shock crossing. This pressure enhancement resulted in the bow shock rapidly retreating Earthward \cite{zou2024}. Additionally, singly charged Helium ions (He$^{+}$) were observed over this time \cite{graham2024}. While He$^{+}$ are rarely seen in the solar wind, He$^{+}$ are known to sometimes occur behind CMEs \cite{gosling1980,schwenn1980,zwickl1982}. 

Throughout this paper we label the shocks by number ($1$--$5$) in the order they are observed. For each shock we calculate the properties, including the shock-normal angle $\theta_{Bn}$, and the Alfv\'{e}n and fast magnetosonic Mach numbers. We note that the upstream plasma conditions can have uncertainties. For ion distributions the angular and energy resolution of FPI is too coarse to well resolve the cold ion beam of the solar wind, which makes the proton densities and temperatures very uncertain. Additionally, ions are sorted by energy per charge and FPI does not resolve different ion species. As a result solar wind He$^{2+}$ will be measured by FPI at higher energies (twice the proton energy when the particle speeds are equal). The proton moments are computed assuming the ion distributions consists of protons alone. As a result significant concentrations of He$^{2+}$ can modify the proton moments. In particular, in the solar wind significant concentrations of He$^{2+}$ will increase the proton density, bulk velocity and temperature moments when integrating over the entire ion distribution. In the solar wind the electron number density can be underestimated due to the core electron population not being fully resolved at low energies. We therefore use the same procedures as in \citeA{graham2024} to calculate the upstream and downstream plasma conditions, viz.: 
\begin{itemize}
\item Bulk proton velocities are determined from the peaks in the one-dimensional (1D) reduced distribution functions to eliminate the contributions from He$^{2+}$ and He$^+$ ions. We observe He$^{2+}$ across all the shocks, and He$^{+}$ is also observed across shocks 1 and 2. The He$^{2+}$ and He$^{+}$ ions are seen in the magnetic cloud at energies twice and four times the proton energy at $\sim 1$~keV, respectively.
\item Upstream $n_e$ is estimated from the electron plasma frequency estimated from Langmuir waves observed upstream of the shocks. Downstream $n_e$ is calculated from flux conservation across the shock. The electron temperatures $T_e$ upstream of the shocks are calculated using $T_e = P_e/(k_B n_e)$, where $P_e$ is the scalar electron pressure calculated from FPI, and $k_B$ is Boltzmann's constant. 
\end{itemize}

To calculate the shock-normal direction $\hat{\bf n}$ and $\theta_{Bn}$ we use the mixed-mode methods \cite{abraham-shrauner1972}. The fast magnetosonic Mach number $M_f$ is calculated assuming an upstream proton temperature of $T_p = 3$~eV. From OMNI data over this interval we estimate $T_p \sim 2 - 4$~eV, so upstream $T_p = 3$~eV is assumed throughout this paper. Table \ref{Tableshockoverview} summarizes the properties of the five bow shocks observed on 24 April 2023. All shocks are quasi-perpendicular $\theta_{Bn} > 45^{\circ}$, with $\theta_{Bn}$ decreasing in time with each shock due to the slow rotation of ${\bf B}$ inside the magnetic cloud. Shocks 1 and 2 are very close to perpendicular shocks. For all shocks the ratio of upstream to downstream ${\bf B}$ is close to $2$ and for all shocks $M_A$ and $M_f$ are below $2$, corresponding to sub-critical shocks \cite{kennel1985}. For the upstream conditions the Alfv\'{e}n speed $V_A$ is much larger than the ion-acoustic speed, resulting in $M_A \approx M_f$. All shocks are characterized by low plasma beta $\beta$. In Table \ref{Tableshockoverview} we calculate the electron plasma beta $\beta_e$; however, since $T_p < T_e$, the total plasma $\beta$ is $\approx \beta_e$, with proton plasma $\beta_p$ of $\sim 0.01$. Aside from $\theta_{Bn}$, the shocks have similar properties, except for shock 2, which was observe during the pressure pulse. This enables us to compare how the shock structure changes with $\theta_{Bn}$ for low Mach number and low $\beta$ conditions. 

\begin{table}
\begin{tabular}{ |p{2.3cm}||p{2cm}|p{2cm}|p{2cm}|p{2cm}|p{2cm}|  }
\hline
Parameter & Shock 1 & Shock 2 & Shock 3 & Shock 4 & Shock 5 \\
\hline
$\theta_{Bn} (^{\circ})$ & $89$ & $87$ & $74$ & $68$ & $66$ \\ 
$V_{sh,n}$ (km~s$^{-1}$) & $50$ & $-140$ & $60$ & $-50$ & $40$ \\ 
$V_{u,n}/V_{d,n}$ & 2.0 & 2.2 & 2.1 & 2.0 & 2.1 \\
$B_d/B_u$ & 2.0 & 1.8 & 2.2 & 2.1 & 2.0 \\
$n_{e,u}$ (cm$^{-3}$) & 5.5 & 9.0 & 6.5 & 6.5 & 6.5 \\
$T_{e,u}$ (eV) & 16 & 11 & 10 & 13 & 10 \\ 
$T_{e,d}/T_{e,u}$ & 6.7 & 3.2 & 12 & 6.0 & 7.8 \\ 
$\beta_e$ & 0.034 & 0.040 & 0.025 & 0.035 & 0.025 \\ 
$M_A$ & 1.8 & 1.4 & 1.9 & 1.6 & 1.7 \\ 
$M_f$ & 1.8 & 1.4 & 1.9 & 1.6 & 1.7 \\ 
\hline
\end{tabular}
\caption{Summary of the properties of the five shocks observed on 24 April 2023. }
\label{Tableshockoverview}
\end{table}

Figures \ref{overviewfig}d--\ref{overviewfig}l provide overviews of shocks 1, 3, and 5 (the inbound bow shock crossings), showing ${\bf B}$ and the evolution of the ion distributions across the shocks. For each shock we rotate the vector data into $(\hat{\bf n},\hat{\bf t}_1,\hat{\bf t}_2)$ coordinates, where $\hat{\bf t}_1$ is aligned with the upstream ${\bf B}_u$ magnetic field and perpendicular to $\hat{\bf n}$, and $\hat{\bf t}_2$ completes the right-hand coordinate system. For each shock $B_{t1}$ increases by approximately a factor of two across the ramp. For shock 1 we observe a shock foot, as indicated by the smaller increase in $B_{t1}$ before the larger rapid increase in $B_{t1}$ at the ramp. For shock 1, $B_{n}$ and $B_{t2}$ remain close to zero across the shock, indicating an approximately perpendicular shock. For shocks 3 and 5, $B_n > 0$ is observed and remains constant across the shocks. We find that $B_{t2}$ remains close to zero, except at the ramp, where we observe a localized peak in $B_{t2} < 0$. There tends to be enhanced fluctuations in $B_{n}$ and $B_{t2}$ across the ramp and in the downstream region for shock 5 compared with shocks 1 and 3. For shocks 1 and 3 we observe quasi-periodic fluctuations in $|{\bf B}|$ in the downstream region. 

Figures \ref{overviewfig}e and \ref{overviewfig}f show the one-dimensional (1D) reduced ion distributions $f_{i}$ along $v_n$ and $v_{t2}$, $f_{i}(v_n)$ and $f_{i}(v_{t2})$, for shock 1. To compute the reduced distributions we have converted ion energy-per-charge to speed using the proton mass $m_p$ and charge $e$. As a result the speeds of He$^{2+}$ and He$^{+}$ will be overestimated by factors of $\sqrt{2}$ and $2$, respectively. In the upstream region we observe the incoming protons at $v_n \approx 500$~km~s$^{-1}$. We also observe small peaks in $f_{i}(v_n)$ at $v_n \approx 700$~km~s$^{-1}$ and $1000$~km~s$^{-1}$, which correspond to He$^{2+}$ and He$^{+}$, respectively. At the shock ramp we observe the upstream proton distribution rapidly decelerate, as seen by the shift in $f_{i}(v_n)$ to lower $v_n$. Downstream of the shock, $f_{i}(v_n)$ and $f_{i}(v_{t2})$ broaden, corresponding to proton heating across the shock. We similarly observe deceleration of He$^{2+}$ and He$^{+}$ across the shock followed by quasi-periodic fluctuations due to the gyromotion He$^{2+}$ and He$^{+}$ downstream of the shock. These quasi-periodic fluctuations are most clearly seen for $f_{i}(v_n)$ (Figure \ref{overviewfig}e). The same behavior was observed for shock 2 and described in detail in \citeA{graham2024}.

We also observe quasi-periodic striations in $f_{i}(v_n)$ and $f_{i}(v_{t2})$, which persist far downstream of the shock. These striations result from a small fraction of the protons being reflected at the ramp, as seen in Figure \ref{overviewfig}e by the increase in $f_i(v_n)$ for $v_n \gtrsim 0$ in the foot region. The reflected protons are accelerated by the solar wind convection ${\bf E}$ and gyro-turned by ${\bf B}$ back to the shock, where they are then transmitted across the ramp. These reflected protons have significantly higher speeds than protons directly transmitted across the ramp, and result in the quasi-periodic striations. 
The He$^{+}$ ions partly obscure part of the striations associated with the reflected protons, and is most clearly in Figure \ref{overviewfig}e for $v_n < 0$. 
Overall, the ion behavior across shock 1, including the presence of He$^{2+}$ and He$^{+}$ is similar to shock 2, which was described in detail in \citeA{graham2024}. 

The 1D reduced ion distributions, $f_{i}(v_n)$ and $f_{i}(v_{t2})$, for shock 3 are shown in Figures \ref{overviewfig}h and \ref{overviewfig}i. Like shock 1 we observe a rapid deceleration along $v_n$ of protons and He$^{2+}$ across the ramp, and a small fraction of reflected protons. Just downstream of the ramp we observe quasi-periodic striations similar to shock 1. However, in contrast to shock 1, these striations quickly broaden after a couple of periods and are not seen after about 04:02:30 UT in Figures \ref{overviewfig}h and \ref{overviewfig}i. Instead, the striations quickly evolve into an energetic halo component of the proton distribution, with only minor variations with time. 

Figures \ref{overviewfig}k and \ref{overviewfig}l show $f_{i}(v_n)$ and $f_{i}(v_{t2})$ across shock 5. Like the previous shocks we observe the deceleration of the protons and He$^{2+}$ across the ramp. For this shock we do not observe any proton reflection at the ramp. As a result we do not observe any striations or a halo associated with energized protons. The transmitted protons across the shock are heated to a similar degree as the directly transmitted protons across shocks 1 and 3. The behavior of protons across shock 4 is very similar to shock 5, namely, reflected protons are not observed. These observations show that the proton behavior varies significantly between the shocks despite the shocks all being quasi-perpendicular and having similar Mach numbers. 

\begin{figure*}[htbp!]
\begin{center}
\includegraphics[width=140mm, height=120mm]{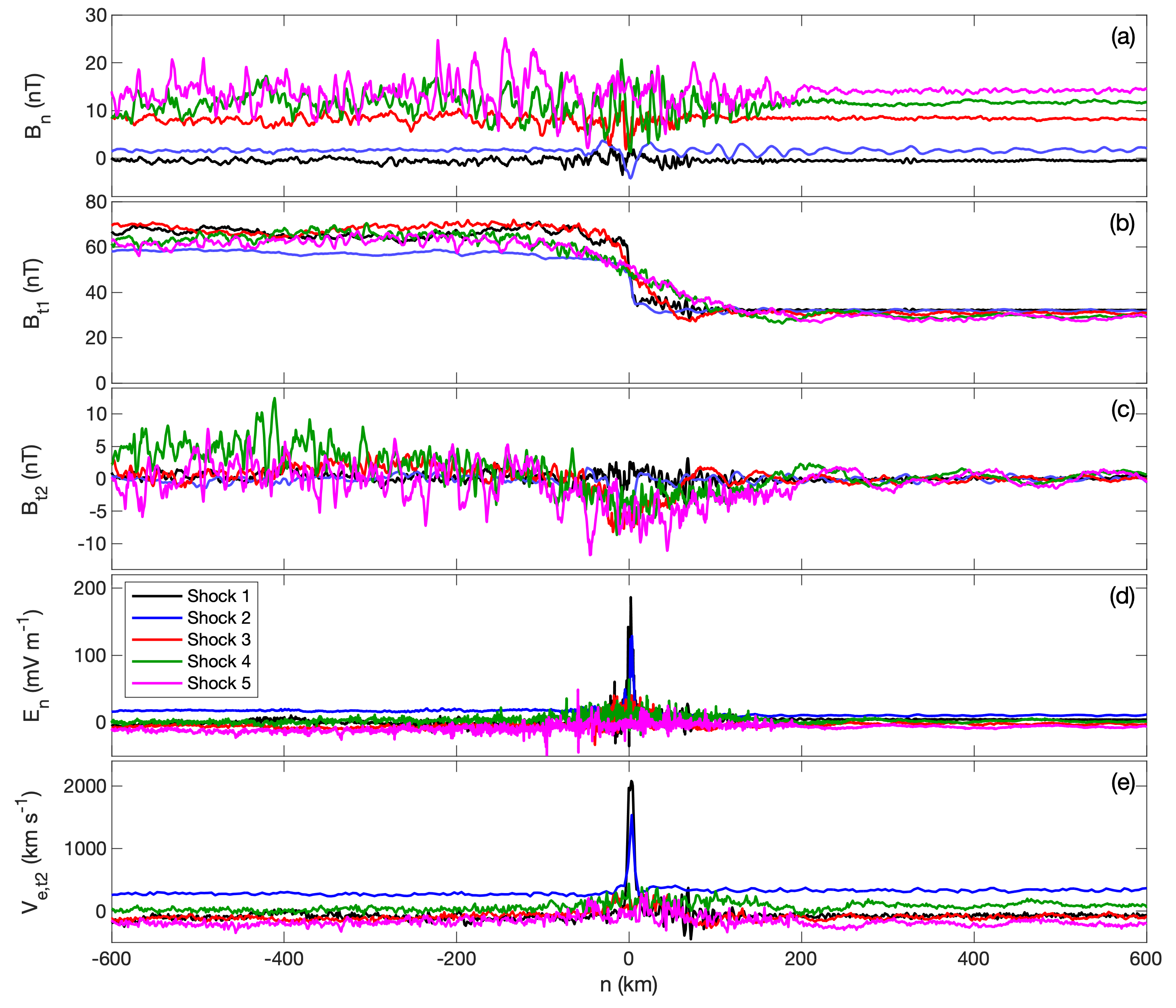}
\caption{Profiles of the five shocks. Time is converted to position using $V_{sh,n}$ with $n = 0$ being centered at the shock ramp. (a) $B_n$, (b) $B_{t1}$, (c) $B_{t2}$, (d) $E_n$, and (e) $V_{e,t2}$. The black, blue, red, green, and magenta curves are shocks 1--5, respectively. $E_n$ and $V_{e,t2}$ are presented in the spacecraft reference frame.}
\label{figfiveshocks}
\end{center}
\end{figure*}

We now investigate how the shock profiles change between the shocks. In Figure \ref{figfiveshocks} we plot the profiles of the five shocks as a function of position along $\hat{\bf n}$. We use the shock speed along $\hat{\bf n}$, estimated from four-spacecraft timing \cite{vogt2011}, to convert time to position and center the shock ramp at $n = 0$~km. We observe the following changes in shock properties: 
\begin{itemize}
\item The ramp width increases with shock number, or equivalently as $\theta_{Bn}$ decreases, as seen by the profiles of $B_{t1}$ in Figure \ref{figfiveshocks}b. Shocks 1 and 2 have extremely narrow ramps, while shocks 4 and 5 have much broader ramps. The average upstream and downstream $B_{t1}$ are similar for each shock, except shock 2 where the downstream $B_{t1}$ is smaller than the other shocks. 
\item The average $B_{n}$ increases with shock number, corresponding to $\theta_{Bn}$ decreasing with time. We observe that the fluctuations in $B_{n}$ and $B_{t2}$ increase with shock number at the ramp and in the downstream region. This is most evident for shocks 4 and 5, which have significantly larger fluctuations in $B_{n}$ and $B_{t2}$ than shocks 1--3. Similarly, enhanced fluctuations in density occur within the ramps of shocks 4 and 5. 
\item For shocks 1 and 2 very large unipolar amplitude $E_n > 100$~mV~m$^{-1}$ occur at the ramp. In contrast, for shocks 3--5, $E_n$ fluctuates at the ramp and a clear unipolar $E_n$ is difficult to observe. Here we have low-pass filtered $E_n$ to minimize the contributions of high-frequency electrostatic waves.
\item For shocks 1 and 2 we observe a very large electron bulk flow $V_{e,t2}$ along $\hat{\bf t}_2$ at the ramp, whereas for shocks 3--5, $V_{e,t2}$ fluctuates at much lower magnitudes. Here, $V_{e,t2}$ serves as a proxy for the current density ${\bf J}$ along $\hat{\bf t}_2$. The peak currents are much stronger in shocks 1 and 2 due to the very narrow ramp. At the shock ramp $E_n$ is primarily supported by the Hall term ${\bf J} \times {\bf B}/(e n_e)$ or equivalently ${\bf E} \approx - {\bf V}_e \times {\bf B}$ \cite{khotyaintsev2023}, so $E_n$ is correlated with $V_{e,t2}$. 
\end{itemize}

In summary, we present an overview of the five bow shock crossings. The shocks are quasi-perpendicular and all have low Mach numbers and low $\beta$. The local shock-normal angles observed by MMS decrease with time, while the ramp width increases. Proton reflection is observed at shocks 1--3, but is absent in shocks 4 and 5. 

\section{Kinetic Ion Behavior Across the Shocks} \label{sec4mod}
In this section we investigate in detail the changes in ion behavior across the five shocks. We compare the observed ion distributions with modelled ion distributions to determine how $\theta_{Bn}$ and the shock width $l$ determine the ion distributions across and downstream of the shock. To model the ion distributions we use Liouville mapping, based on the observed upstream and downstream parameters of the shocks. The Liouville mapping method is described in \ref{app1}. We model the changes in $B_{t1}$, $n_e$, and $P_e$ across the shocks using hyperbolic tangent functions, with characteristic width $l$ [equations (\ref{Bt1model})--(\ref{Pemodel})]. The model ${\bf E}$ is calculated using Ohm's law. 

\subsection{Comparison with Observations}
For the numerical model we use upstream and downstream parameters based on MMS observations of the shocks. Table \ref{Tableshockprops} summarizes the parameters used in the model. For each of the shocks the compression ratio of ${B_{t1}}$ and $n_e$ is close to $2$ and the upstream $B_{t1}$ is relatively constant. The changes in $V_{u,n}$ in the shock frame are primarily due to the inward and outward motion of the bow shock across the spacecraft. To estimate the shock width $l$ we fit a hyperbolic tangent function [equation (\ref{Bt1model})] to the observed $B_{t1}$ by converting time to position using the estimated shock speed $V_{sh,n}$ in the spacecraft frame. We find that $l$ increases with shock number for the five shocks, and is correlated with the increase in $B_{n}$, or equivalently the decrease in $\theta_{Bn}$, as seen in Figure \ref{figfiveshocks}. In Table \ref{Tableshockprops} we provide the upstream proton inertial length $d_p = c/\omega_{pp}$ and proton convective gyroradius $V_{u,n}/\Omega_{cp}$, where $\omega_{pp}$ and $\Omega_{cp}$ are the angular proton plasma and cyclotron frequencies computed from the upstream conditions. For shocks 1--3 we find that $l$ is well below $d_p$ and $V_{u,n}/\Omega_{cp}$, while for shocks 4 and 5, $l$ is comparable to these length scales. Thus, for shocks 4 and 5 the protons crossing the ramp cannot be assumed to be completely unmagnetized. 

For subcritical shocks a theoretical prediction for the shock width is given by \cite{moiseev1963,mellott1984}
\begin{equation}
l_{sh} \approx \frac{2 \pi d_p \cos{\theta_{Bn}}}{\sqrt{M_A^2 - 1}}.
\label{lengthpred}
\end{equation}
Equation (\ref{lengthpred}) is based on the low Mach number shock behaving like a standing magnetosonic/whistler wave. The estimates of $l_{sh}$ are shown in Table \ref{Tableshockprops}, and are somewhat lower but similar to $l$, with $l_{sh}$ tending to increase as $\theta_{Bn}$ decreases. This comparison suggests that for these shocks the width is determined primarily by $\theta_{Bn}$. 

For each shock the shock potential $\phi$ in the normal incidence frame (NIF) is estimated from the upstream and downstream conditions using equation (\ref{phieq}). The predicted $\phi$ is close to $1$~kV, except shock 2, which has $\phi \approx 500$~V. The predicted $\phi$ is independent of $l$. However, the peak amplitude of the normal electric field $E_{n,max}$ in the NIF calculated using equation (\ref{Enmodel}) is primarily determined by $l$ and decreases substantially between shocks 1 and 5, as $l$ increases. For the shocks considered here the electron pressure gradient contributes $\approx 10 - 20$~\% to $E_n$ and $\phi$. Finally, we calculate the upstream proton kinetic energy, $E_{sw} = m_p V_n^2/2$. For each shock we find that $e \phi$ is below but comparable to $E_{sw}$, suggesting that $\phi$ can be sufficiently large to specularly reflect some of the incoming protons, consistent with observation of reflected protons for shocks 1--3. Similarly, large $\phi$ have been reported in observations for relatively low $M_A$ shocks \cite{dimmock2012}. 

\begin{table}
\begin{tabular}{ |p{2.8cm}||p{1.8cm}|p{1.8cm}|p{1.8cm}|p{1.8cm}|p{1.8cm}|  }
\hline
Parameter & Shock 1 & Shock 2 & Shock 3 & Shock 4 & Shock 5 \\
\hline
$B_{t1,u}$ (nT) & 32 & 32 & 30 & 30 & 29 \\ 
$B_{t1,d}$ (nT) & 70 & 59 & 70 & 65 & 63 \\ 
$B_{n}$ (nT) & -0.35 & 1.7 & 8.6 & 12 & 13.5 \\
$n_u$ (cm$^{-3}$) & 5.5 & 9 & 6.5 & 6.5 & 6.5 \\
$n_d$ (cm$^{-3}$) & 11 & 17 & 13.5 & 13 & 13.5 \\
$P_{e,u}$ (nPa) & 0.014 & 0.015 & 0.01 & 0.014 & 0.01 \\ 
$P_{e,d}$ (nPa)& 0.19 & 0.10 & 0.35 & 0.22 & 0.16 \\ 
$V_u$ (km~s$^{-1}$) & -540 & -320 & -520 & -430 & -480 \\ 
$l$ (km) & 5 & 5 & 25 & 80 & 100 \\ 
$l_{sh}$ (km) & 5 & 30 & 60 & 140 & 120 \\ 
$d_{p}$ (km) & 100 & 80 & 90 & 90 & 90 \\ 
$|V_{u,n}|/\Omega_{cp}$ (km) & 180 & 100 & 170 & 140 & 160 \\ 
$E_{\mathrm{n,max}}$ (mV~m$^{-1}$) & 130 & 51 & 24 & 6 & 4 \\ 
$\phi$ (V) & 1300 & 510 & 1200 & 980 & 870 \\ 
$E_{sw}$ (eV) & 1500 & 530 & 1400 & 970 & 1200 \\ 
\hline
\end{tabular}
\caption{Properties of the bow shock crossing used to model the ion distributions. }
\label{Tableshockprops}
\end{table}

\begin{figure*}[htbp!]
\begin{center}
\includegraphics[width=160mm, height=210mm]{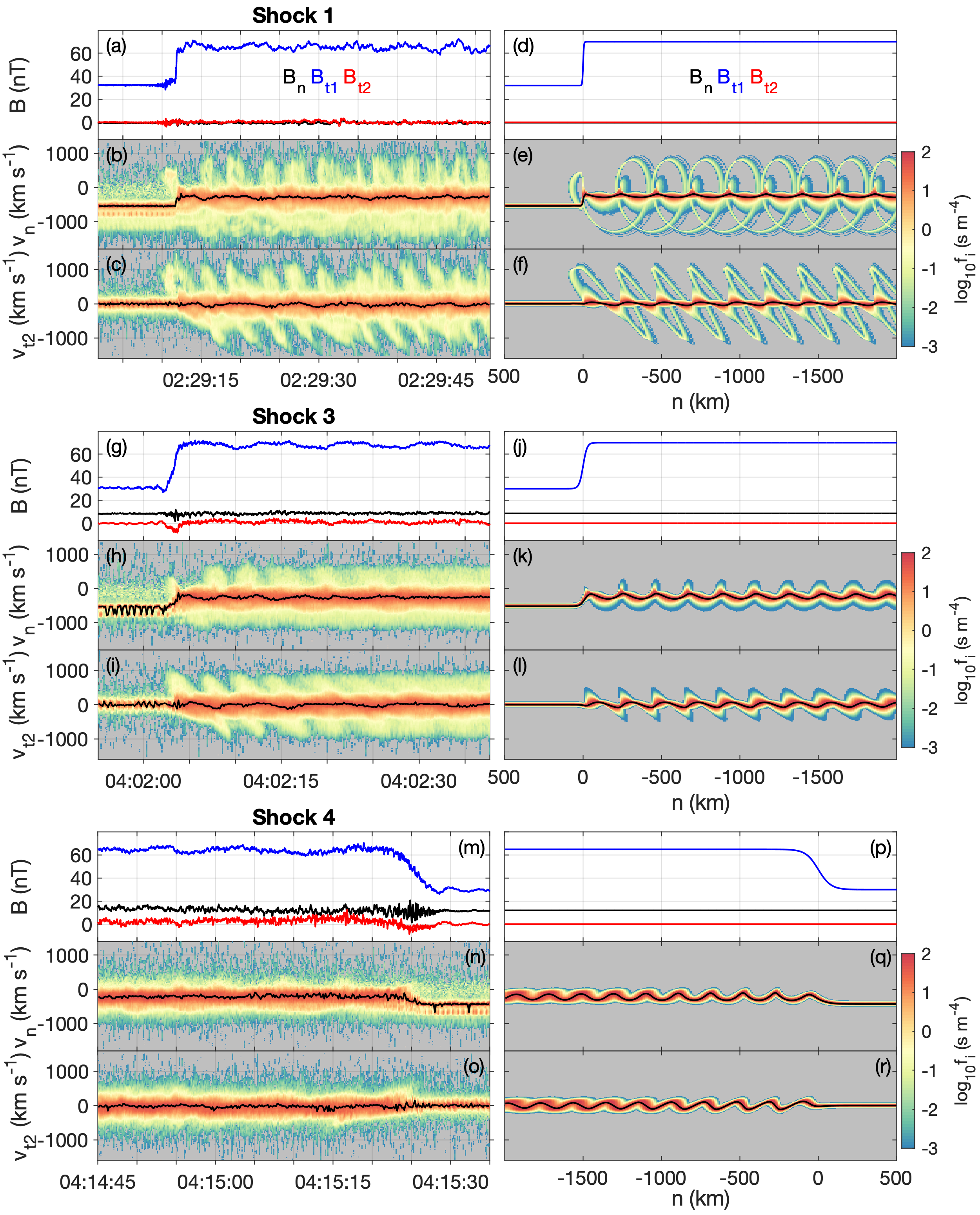}
\caption{Comparison of observed ion distributions across shocks 1, 3, and 4 with the modelled distributions. Panels (a)--(f) show shock 1, Panels (g)--(l) show shock 3, 
and Panels (m)--(r) show shock 3. (a) ${\bf B}$ in $(\hat{\bf n},\hat{\bf t}_1,\hat{\bf t}_2)$ coordinates, (b) and (c) 1D reduced ion distributions along $v_n$ and $v_{t2}$. 
(d)--(f) Same quantities as (a)--(c) obtained from the numerical model. Panels (g)--(j) and (m)--(r) are in the same format as panels (a)--(f). All distributions are presented in the NIF. The black lines in the observed reduced distribution plots are $v_n$ and $v_{t2}$ corresponding to the peaks in $f(v_n)$, and $f(v_{t2})$, which we use as a proxy for the bulk velocities. The black lines in the modelled distributions correspond to the bulk velocity.}
\label{modelshockcomp}
\end{center}
\end{figure*}

Using the parameters in Table \ref{Tableshockprops} we calculate the proton distributions across the five shocks using the Liouville mapping routine detailed in \ref{app1}. In Figure \ref{modelshockcomp} we plot the observed and modelled ${\bf B}$, $f_i(v_n)$, and $f_i(v_{t2})$ for shocks 1, 3, and 4, so the observed and modelled distributions can be directly compared. For the observed shocks the time interval corresponds to the same spatial domain used in the model assuming constant $V_{sh,n}$. 

In Figures \ref{modelshockcomp}a--\ref{modelshockcomp}f we compare the observed ion distributions with the model proton distributions for shock 1. We observe a small fraction of reflected protons in both observations and the model distributions. These protons correspond to the slowest moving protons along $v_n$ at the ramp and are reflected by $\phi$. The majority of protons are decelerated by $\phi$ and transmitted across the shock without reflection due to $e \phi < E_{sw}$. Both the reflected and transmitted protons undergo periodic fluctuations downstream of the ramp. In Figure \ref{modelshockcomp}e the reflected protons form loop-shaped structures along $v_n$. In Figure \ref{modelshockcomp}b the reflected protons have similar speeds although the loop-like structures are difficult to resolve. Along $v_{t2}$, similar striations occur in both the observations and model for the reflected protons. The fact that comparable phase-space densities of reflected protons in the model and observations suggests that equation (\ref{phieq}) provides a reasonable estimate of $\phi$ for this shock. The same behavior is observed for shock 2 \cite{graham2024}. 

We observe some evidence of quasi-periodic motion in the transmitted protons due to gyrophase bunching of the protons \cite{gedalin1996}. The black lines in Figures \ref{modelshockcomp}b and \ref{modelshockcomp}b are $v_n$ and $v_{t2}$ corresponding to the peaks in $f_i(v_n)$ and $f_i(v_{t2})$. We use this as a proxy for the proton bulk velocity ${\bf V}_p$ because the proton moments calculated from FPI are affected by He$^{2+}$. At the ramp the protons are rapidly decelerated by $\phi$ and no longer move at the ${\bf E} \times {\bf B}$-drift velocity. Instead the entire distribution fluctuations around the ${\bf E} \times {\bf B}$-drift velocity, as seen in Figures \ref{modelshockcomp}b--\ref{modelshockcomp}c and \ref{modelshockcomp}e--\ref{modelshockcomp}f. The black lines in Figures \ref{modelshockcomp}e and \ref{modelshockcomp}f indicate ${\bf V}_p$ calculated from the model distributions. These results suggest that gyrophase bunching of protons downstream of the shock is occurring. 

Figures \ref{modelshockcomp}g--\ref{modelshockcomp}l show the comparison of the model and observed distributions for shock 3. In the model we observe a smaller fraction of reflected protons compared with the observations, which might suggest that $\phi$ is underestimated in the model. In the model we observe periodic fluctuations in proton distributions downstream of the shock, due to gyrophase bunching. These fluctuations are seen in observations in Figures \ref{modelshockcomp}h and \ref{modelshockcomp}i for both the transmitted the reflected protons. In the model we find that there is a gradual broadening of the distributions along $v_n$ and $v_{t2}$, which corresponds to gyrophase mixing of the transmitted protons, and corresponds to a gradual increase in temperature of the distribution. This gyrophase mixing is likely responsible for the smoothing out of the structured distributions associated with the reflected protons in the observations. In observations this gyrophase mixing appears to occur more rapidly than in the model. Overall, the modelled and observed distributions exhibit similar characteristics, although the model underestimated the fraction of reflected protons. We note that the fraction of reflected protons depends strongly on the upstream bulk speed, upstream $T_p$, $\phi$, and $l$, so relatively small changes in these parameters could strongly affect the fraction of reflected protons. 

Figures \ref{modelshockcomp}g--\ref{modelshockcomp}l show the comparison of the model and observed distributions for shock 4. In observations we do not observe proton reflection, similar to shock 5. Similarly, in the model we do not observe any reflected protons, despite $E_{sw} \approx e \phi$. In observations and the model, the protons are decelerated across the ramp, although over a longer interval than shocks 1 and 3, due to the increase in $l$. In the model the downstream protons undergo periodic fluctuations along $v_n$ and $v_{t2}$ associated with gyrophase bunching. However, in observations we do not observe any clear evidence of gyrophase bunching of the protons. In observations this lack of gyrophase bunching may result from scattering due to the enhanced magnetic field fluctuations across then ramp, compared with shocks 1 and 3. The same behavior occurs in the observed and modelled distributions of shock 5. This indicates that the Liouville mapping model lacks some processes that modify the downstream proton distributions. 

In each of the shock crossings we also observe gyrophase bunching of He$^{2+}$ in the downstream region. In computing $f_i$ we assumed the proton mass and charge when converting energy-per-charge to speed. The relation between the velocity of ions seen by FPI in the reduced distributions to the ion velocity in the NIF is given by \cite{graham2024}
\begin{equation}
{\bf v}_{sc} = \sqrt{\frac{m_i}{Z m_p}} \left( {\bf v}_i + {\bf V}_{sh} \right),
\label{vioneq}
\end{equation}
where ${\bf V}_{sh}$ is the speed of the NIF with respect to the spacecraft. A consequence of equation (\ref{vioneq}) is that proton and He$^{2+}$ populations tend to be shifted apart from each other in the spacecraft frame. For the shocks considered here we observe quasi-periodic peaks in $f_i(v_n)$ associated with He$^{2+}$ for $v_n < 0$, where the observed $v_{sc,n}$ of He$^{2+}$ exceeds the proton speeds. These peaks are observed for all five shocks and are seen in Figures \ref{modelshockcomp}b, \ref{modelshockcomp}h, and \ref{modelshockcomp}n, and indicate quasi-periodic gyromotion and gyrophase bunching of He$^{2+}$ downstream of the shocks \cite{gedalin2017}. 

Overall, we find good agreement between the model and observed proton distributions. For shock 1, we observe reflected protons in observations and the model, which produce periodic striations that persist far downstream of the shock. For shock 3 we observe reflected protons, producing quasi-periodic striations behind the ramp, which are quickly smoothed out to form a halo distribution. For shock 4 no proton reflection occurs despite $e \phi \approx E_{sw}$, consistent with the model predictions. To understand why these differences in proton behavior occur we investigate the dependence on shock width $l$ and shock-normal angle $\theta_{Bn}$. 

\subsection{Dependence on Shock Width}
\subsubsection{Numerical Results}
\begin{figure*}[htbp!]
\begin{center}
\includegraphics[width=140mm, height=160mm]{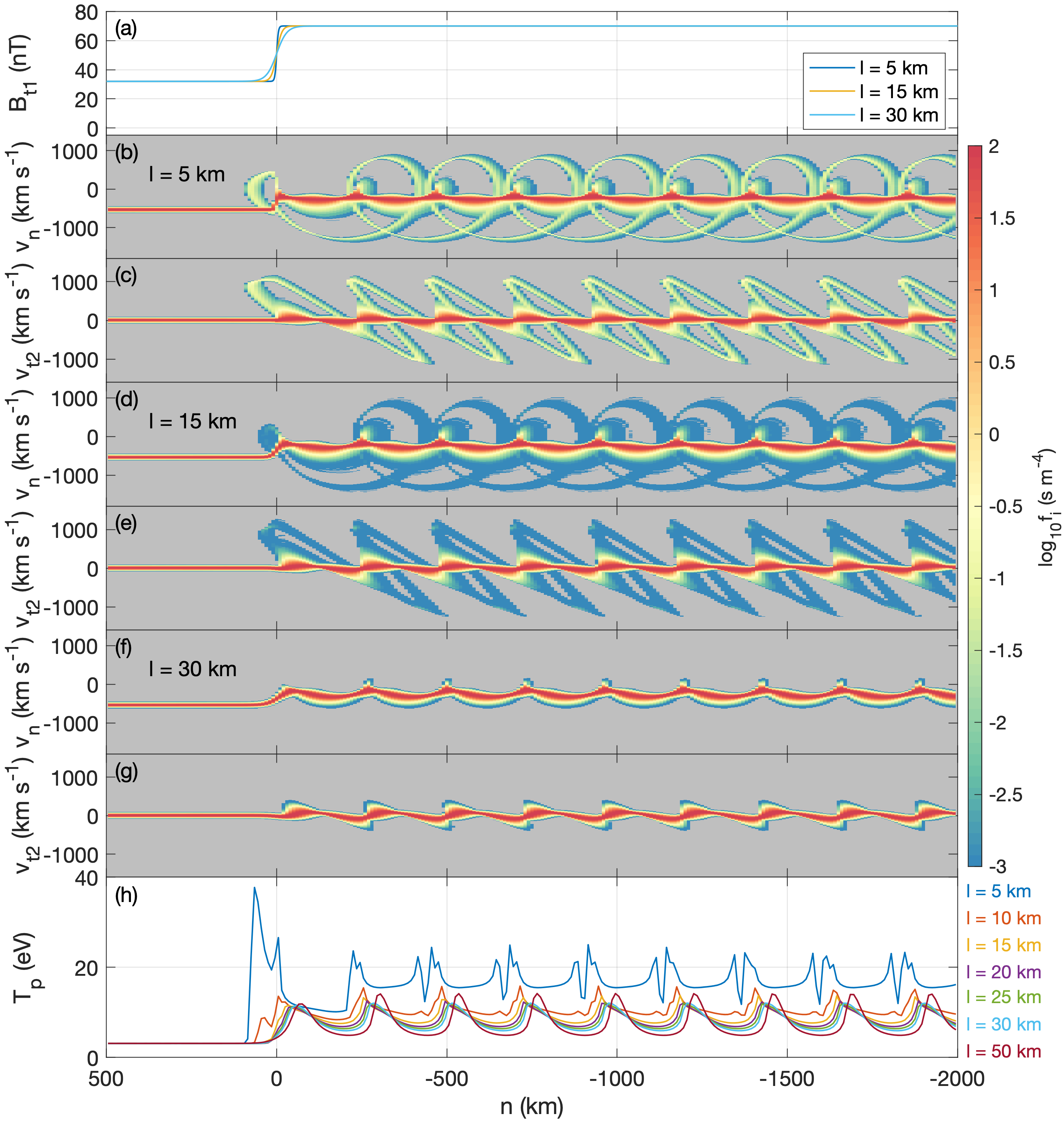}
\caption{Dependence of proton distributions across the shock on shock width $l$. For the model shock parameters we use 
shock 1 conditions with $B_n = 0$~nT, corresponding to a perpendicular shock. (a) Profiles of $B_{t1}$ for $l = 5$~km (dark blue), $l = 15$~km (yellow), and $l = 30$~km (light blue). (b) and (c) Reduced 1D proton distributions along $v_n$ and $v_{t2}$ for $l = 5$~km. (d)--(e) and (f)--(g) Same format as (b)--(c) for $l = 15$~km and $l = 30$~km, respectively. 
(h) Scalar proton temperature $T_p$ across the shock for $l = 5$, $10$, $15$, $20$, $25$, $30$, and $50$~km (dark blue, red, yellow, purple, green, light blue, and maroon lines, respectively).}
\label{modellfig}
\end{center}
\end{figure*}

Using the numerical model we investigate the change in proton behavior across the shock as a function of $l$. To illustrate the dependence on $l$ we use shock 1 conditions with $B_n = 0$, corresponding to a perpendicular shock, and vary $l$. Figure \ref{modellfig} shows the modelled $f_i(v_n)$ and $f_i(v_{t2})$ for shocks with $l = 5$~km, $l = 15$~km, and $l = 30$~km. For $l = 5$~km the distributions are almost identical to the shock 1 model. However, for $l = 15$~km the fraction of reflected protons is substantially reduced. In particular, the values of $f_i$ associated with the reflected protons are over an order of magnitude smaller than for $l = 5$~km. When $l$ is increased further we no longer observe any reflected protons. Figures \ref{modellfig}f and \ref{modellfig}g show that for $l = 30$~km no reflected protons occur. As $l$ is further increased no proton reflection is observed. 

The shock width can therefore significantly affect the total downstream proton temperatures $T_p$. Figure \ref{modellfig}h shows the scalar $T_p$ across the model shocks with $l$ ranging from $5$~km to $50$~km. For $l = 5$~km, $T_p$ is substantially larger than for the cases with $l \geq 10$~km due to the contribution of the reflected protons. Even at $l = 10$~km the fraction of reflected protons is substantially reduced and only results in a small increase in $T_p$. In all cases the downstream $T_p$ exhibits periodic fluctuations along $\hat{\bf n}$. For $l \gtrsim 20$~km, when proton reflection is negligible there are only small differences in $T_p$ as $l$ is further increased. Namely, as $l$ increases the minima in $T_p$ decreases and the relative position of peaks in $T_p$ are shifted further downstream. The periodic fluctuations in $T_p$ are due to gyrophase bunching of the protons \cite{gedalin1996}. 

\subsubsection{Theoretical Explanation}
To understand why the fraction of reflected protons depends on $l$ we consider the equations of motion [equation (\ref{eqsofmotion})] for a perpendicular shock, which are given by: 
\begin{equation}
\frac{d v_n}{d t} = \frac{e}{m_p} \left( E_n - B_{t1} v_{t2} \right), 
\label{eqvn1}
\end{equation}
\begin{equation}
\frac{d v_{t2}}{d t} = \frac{e}{m_p} \left( E_{t2} + B_{t1} v_{n} \right), 
\label{eqvt21}
\end{equation}
in the NIF. In the limit $l \rightarrow 0$, the change in $v_n$ is given by
\begin{equation}
\frac{1}{2} m_p \left( v_{u,n}^2 - v_{d,n}^2 \right) = e \phi.
\label{eqdeltavn}
\end{equation}
For $1/2 m_p v_{u,n}^2 < e \phi$, protons are expected to be specularly reflected. 
However, for a finite $l$, $v_{t2} \neq 0$ can develop within the ramp, which will contribute to the acceleration along $\hat{\bf n}$ according to equation (\ref{eqvn1}). As the protons approach the ramp $v_{n}$ decreases and $B_{t1}$ increases resulting in an initial $d v_{t2}/dt < 0$, which further decelerates incoming protons along $\hat{\bf n}$. However, based on the model bulk velocity ${\bf V}$ moments this $v_{t2}$ contribution is relatively small and at a distance of $\approx l$ behind the center of the ramp $V_{t2}$ becomes positive, which counters the effect of $E_n$ and $\phi$. This $V_{t2} > 0$ substantially exceeds the initial $V_{t2} < 0$. Further downstream $V_{t2}$ undergoes periodic fluctuations in $V_{t2}$. We also find that the oscillations in $V_{t2}$ increase in magnitude with $l$. The effect of $v_{t2}$ in equation (\ref{eqvn1}) is then to reduce the deceleration in $v_n$ due to $\phi$, reducing the proportion of reflected protons. Moreover, as $l$ increases the reflection point will be moved further downstream of the center of the ramp, so even if $v_n > 0$ occurs for a wider shock, the protons are less likely to reach upstream region and be significantly accelerated along $\hat{\bf t}_{2}$. 

In brief, we find that as $l$ increases, the fraction of reflected protons decreases even for $l \ll d_p$. Increasing $l$ can eliminate proton reflection entirely, even though proton reflection is predicted according to equation (\ref{eqdeltavn}). This accounts for the lack of proton reflection observed in shocks 4 and 5, with $l \sim d_p$, despite the predicted $\phi$ being sufficiently large to reflect some of the incoming protons. The reduction in proton reflection is due to the variations in $v_{t2}$ across the ramp and the reflection point moving downstream of the center of the ramp as $l$ increases. The shock width can thus have a significant effect on the total downstream proton temperature and thermal pressure. 

\subsection{Dependence on Shock Normal Angle}
\subsubsection{Numerical Results}
\begin{figure*}[htbp!]
\begin{center}
\includegraphics[width=140mm, height=160mm]{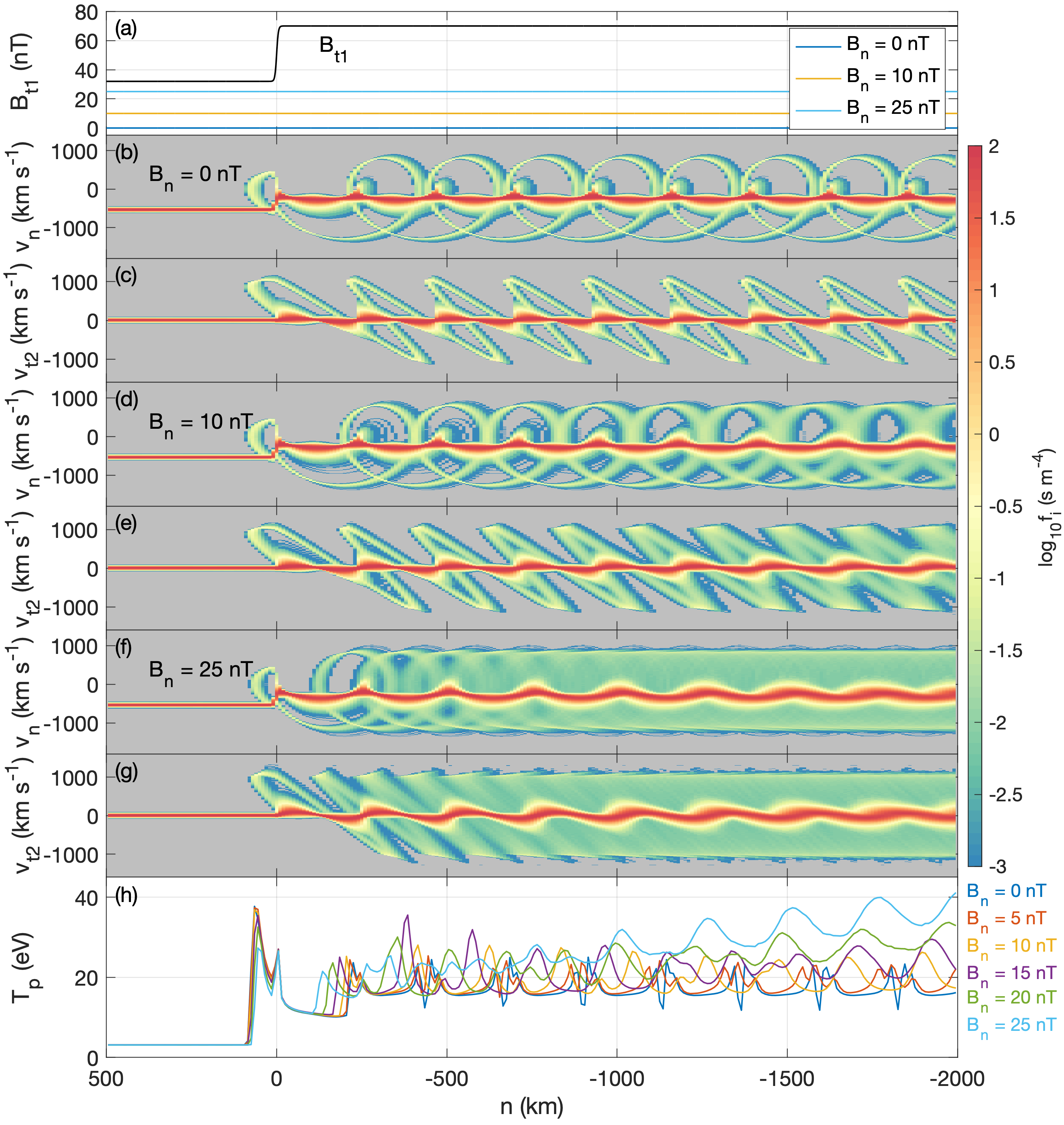}
\caption{Dependence of ion distributions across the shock on $\theta_{Bn}$. For the model shock parameters we use shock 1 conditions and variable $B_n$. (a) Profiles and $B_{t1}$ and $B_n = 0$~nT (dark blue), $B_n = 10$~nT (yellow), and $B_n = 25$~nT (light blue). (b) and (c) Reduced 1D proton distributions along $v_n$ and $v_{t2}$ for $B_n = 0$~nT. (d)--(e) and (f)--(g) Same format as (b)--(c) for $B_n = 10$~nT and $B_n = 25$~nT, respectively. (h) Scalar proton temperatures $T_p$ across shocks with $B_n = 0$, $5$, $10$, $15$, $20$, and $25$~nT (dark blue, red, yellow, purple, green, and light blue lines, respectively).}
\label{modelBnfig}
\end{center}
\end{figure*}

We now consider the effect of $\theta_{Bn}$ on the proton distributions across the shock by varying $B_n$. To illustrate the dependence on $\theta_{Bn}$ we use shock 1 conditions and vary $B_n$. Figure \ref{modellfig} shows the modelled $f_i(v_n)$ and $f_i(v_{t2})$ for shocks with $B_n = 0$~nT, $B_n = 10$~nT, and $B_n = 25$~nT, corresponding to $\theta_{Bn} = 90^{\circ}$, $73^{\circ}$, and $52^{\circ}$, respectively. In each case we observe a similar fraction of reflected protons, so the change in $B_n$ here does not significantly affect proton reflection. For $B_n = 0$~nT  we observe periodic fluctuations in the transmitted and reflected proton populations, as described above, which are independent of distance along $\hat{\bf n}$ downstream of the shock (Figures \ref{modelBnfig}b and \ref{modelBnfig}c). For $B_n = 10$~nT we observe very similar behavior close to the ramp (Figures \ref{modelBnfig}d and \ref{modelBnfig}e). Further downstream of the ramp there is a broadening of the distributions associated with the reflected protons. For $B_n = 25$~nT the distributions of reflected protons quickly broaden to form an approximately uniform energetic distribution, which does not vary with distance (Figures \ref{modelBnfig}f and \ref{modelBnfig}g). 
%The reflected protons appear as an energetic halo population in the 1D reduced distributions and as a ring in the 2D reduced distributions in the $v_n$--$v_{t2}$ plane. 

For $\theta_{Bn} = 90^{\circ}$ the reflected protons downstream of the shock are gyrophase bunched, meaning they occupy very localized regions in the $v_{n}-v_{t2}$ plane at a given position along $n$, which results in the periodic structures in $f_i(v_n)$ and $f_i(v_{t2})$. Gyrophase bunching also occurs for the directly transmitted protons. For $\theta_{Bn} < 90^{\circ}$ the localized regions in the $v_{n}-v_{t2}$ plane where reflected protons occur broadens with distance downstream of the ramp. This broadening occurs more rapidly as $\theta_{Bn}$ decreases, or equivalently $B_n$ increases. Eventually, the reflected protons form a ring distribution in the $v_{n}-v_{t2}$ plane. The same process occurs for transmitted protons, which form ring-like distributions at much lower speeds. The gyrophase mixing for reflected protons develops over a much shorter distance behind the ramp compared with the transmitted protons (Figures \ref{modelBnfig}f and \ref{modelBnfig}g). 

%For oblique shocks, the equation of motion along $\hat{\bf t}_1$ is given by
%\begin{equation}
%\frac{d v_{t1}}{d t} = \frac{e}{m_p} \left( E_{t1} + v_{t2} B_n \right),
%\label{eqmotionvt1}
%\end{equation}
%in the NIF. For incoming protons $E_{t1}$ associated with the Hall electric field will accelerated to $v_{t1} < 0$. For reflected protons $v_{t2} > 0$ after reflection due to due to acceleration by the upstream convection field. This results in $v_{t1} > 0$ when the reflected protons cross the ramp. Along $v_{t1}$ we observe similar gyrophase mixing of the reflected protons and $v_{t1} < 0$ is typically satisfied. 

%For $\theta_{Bn} = 90^{\circ}$ the directly transmitted protons undergo periodic fluctuations due to gyrophase bunching. As $\theta_{Bn}$ decreases gyrophase mixing occurs, resulting in the proton distributions broadening as distance increases from the ramp. This broadening occurs more rapidly $\theta_{Bn}$ decreases, and corresponds to a faster increase in the temperature of transmitted protons. In Figures \ref{modelBnfig}f and \ref{modelBnfig}g, the gyrophase mixing of reflected protons occurs more rapidly than for the directly transmitted protons. 

This increased gyrophase mixing with increasing $B_n$ results in the scalar proton temperature $T_p$ increasing more rapidly downstream of the shock. Figure \ref{modelBnfig}h shows $T_p$ for $0$~nT~$ < B_n < 25$~nT. In all cases there is an initial increase in $T_p$ just ahead of the ramp due to the reflected protons. Just downstream of the ramp $T_p$ undergoes periodic fluctuations for small $B_n$. As $B_n$ increases, $T_p$ increases more rapidly, due to the more rapid gyrophase mixing of the reflected and transmitted protons. Additionally, for small $B_n$, $dv_{t1}/dt \approx 0$, so heating along this direction is negligible. As $B_n$ increases, proton acceleration along $\hat{\bf t}_1$ increases, resulting in proton heating along $\hat{\bf t}_1$. 

\subsubsection{Theoretical Explanation}
We now consider in detail why gyrophase mixing depends on $\theta_{Bn}$ and occurs more rapidly for reflected protons compared with transmitted protons. For $\theta_{Bn} = 90^{\circ}$ the equation of motion are simple in the downstream region due to $d v_{t1}/dt = 0$. As a result all protons fluctuate about the ${\bf E} \times {\bf B}$ velocity. In the downstream region the proton velocities along $\hat{\bf n}$ as a function time are given by: 
\begin{equation}
v_n(t) = - \frac{E_{t2}}{B_{t1}} + \left( v_{n,0} + \frac{E_{t2}}{B_{t1}} \right) \cos{\left( \Omega_{cp} t \right)},
\label{vnperp}
\end{equation}
where $v_{n,0}$ is assumed to be the initial speed just behind the ramp. From equation (\ref{vnperp}) the wavelength $l_p$ of the oscillations is then given by \cite{balikhin2008,ofman2009}: 
\begin{equation}
l_p = \frac{2 \pi E_{t2}}{B_{t1} \Omega_{cp}} = \frac{2 \pi |V_{d,n}|}{\Omega_{cp}},
\label{lp1}
\end{equation}
where $|V_{n,d}|$ is the average downstream speed along $\hat{\bf n}$, and $l_p$ is defined to be positive. Since equation (\ref{lp1}) is independent of the initial speed just behind the ramp, the wavelength of all protons is the same, and the periodic oscillations are predicted to persist far downstream of the shock without gyrophase mixing for both transmitted and reflected protons. This explains the behavior of reflected protons downstream of shocks 1 and 2, which are nearly perpendicular shocks. 

In the downstream region of an oblique shock the relevant equations of motion can be written as 
\begin{equation}
v_n(t) = - \frac{B_{t1}}{B_n} v_{t1}(t) + v_c,
\label{vnobl1}
\end{equation}
\begin{equation}
\frac{d^2 v_n}{d t^2} + \Omega_{cp}^2 v_n = \frac{e^2}{m_p^2} \left( B_n^2 v_c - B_{t1} E_{t2} \right), 
\label{vnobl2}
\end{equation}
where $v_c$ is a speed resulting from the constant of integration to obtain equation (\ref{vnobl1}). Equations (\ref{vnobl1}) and (\ref{vnobl2}) show that for oblique shocks protons will undergo periodic fluctuations along $v_{t1}$, as well as along $v_n$. From equation (\ref{vnobl2}) we obtain
\begin{equation}
v_n(t) = \frac{B_n^2 v_c - B_{t1} E_{t2}}{B_d^2} + v_1 \cos{\left( \Omega_{cp} t \right)},
\label{vnobl3}
\end{equation}
where $v_1$ is a speed determined by the initial conditions just downstream of the ramp. 
From equation (\ref{vnobl3}) the oscillation wavelength is given by
\begin{equation}
l_p = \frac{2 \pi \left( B_{t1,d} E_{t2} - B_n^2 v_c \right)}{B_d^2 \Omega_{cp,d}}. 
\label{lp2}
\end{equation}
For $B_n \rightarrow 0$, equation (\ref{lp2}) reduces to equation (\ref{lp1}). Equation (\ref{lp2}) shows that for finite $B_n$, $l_p$ can increase or decrease depending on the sign of $v_c$, which is determined by equation (\ref{vnobl1}). For the wavelength $l_{p,t}$ of the transmitted protons we find that $l_{p,t} \approx l_p$ from equations (\ref{vnobl1}) and (\ref{lp2}). However, Figure \ref{modelBnfig} shows that there is a very small increase in $l_{p,t}$ as $B_n$ increases. 

%For transmitted protons we can estimate $v_c$ from the average $V_{n,d}$ and $V_{t1,d}$. From the Rankine-Hugoniot jump conditions, $V_{t1,d}$ is given by 
%\begin{equation}
%V_{t1,d} = V_{n,d} \frac{\left( B_{t1,d} - \frac{n_d}{n_u} B_{t1,u} \right)}{B_n}.
%\label{Vt1deq}
%\end{equation}
%Substituting equations (\ref{vnobl1}) and (\ref{Vt1deq}) into equation (\ref{lp2}) yields
%\begin{equation}
%l_{p,t} = \frac{2 \pi |V_{n,d}|}{\Omega_{cp}} \left( 1 + \frac{B_{t1,d}^2}{B_d^2} - \frac{B_{t1,d} B_{t1,u}}{B_d^2} \frac{n_d}{n_u} \right). 
%\label{lp3}
%\end{equation}
%For the parameters used in Figure \ref{modelBnfig} we obtain $l_p = 250$~km and $l_{p,t} = 260$~km for the cases where $B_n = 0$~nT and $B_n = 25$~nT, respectively. This slight increase in $l_{p,t}$ is consistent with the model and can be seen by comparing Figures \ref{modelBnfig}b--\ref{modelBnfig}g far downstream of the ramp. As $\theta_{Bn}$ approaches $90^{\circ}$, equation (\ref{lp3}) becomes unreliable, so for nearly perpendicular shocks, equation (\ref{lp1}) should be used instead. 

To calculate the wavelength of the fluctuations associated with reflected protons we estimate $v_c$ just downstream of the ramp assuming specular reflection. For $B_n > 0$~nT, reflected protons will be accelerated along $\hat{\bf t}_1$ and $\hat{\bf t}_2$. By assuming the reflected protons remain in the upstream region for half a gyroperiod, the speed along $\hat{\bf t}_1$ when the proton returns to the ramp is 
\begin{equation}
V_{t1} \approx - \frac{2 B_n B_{t1,u} V_{u,n}}{B_u^2}.
\label{Vt1eq2}
\end{equation}
We also assume $V_n \approx V_{u,n}$ for reflected protons crossing the ramp (cf.~Figure \ref{modelBnfig}). From equations (\ref{vnobl1}), (\ref{lp2}), and (\ref{Vt1eq2}) we obtain
\begin{equation}
l_{p,r} \approx \frac{2 \pi |V_{u,n}|}{B_d^2 \Omega_{cp,d}} \left( \frac{n_u B_{t1,d}^2}{n_d} + B_n^2 \left[ 1 - \frac{2 B_{t1,d} B_{t1,u}}{B_u^2} \right] \right).
\label{lp4}
\end{equation}
Equation (\ref{lp4}) reduces to equation (\ref{lp1}) for $B_n = 0$~nT. 
We note that equation (\ref{lp4}) is idealized and neglects the effect of $E_{t1}$ due to the Hall field and finite $l$ of the ramp, so equation (\ref{lp4}) likely underestimates $l_{p,r}$ as $B_n$ becomes large. Equation (\ref{lp4}) predicts that $l_{p,r}$ will decrease as $B_n$ increases. As a result the transmitted and reflected protons will have distinct spatial scales. This is most clearly seen in Figures \ref{modelBnfig}f and \ref{modelBnfig}g just behind the ramp. 
For the model parameters used in Figure \ref{modelBnfig}, equation (\ref{lp4}) predicts $l_{p,r} = 220$~km and $120$~km for $B_n = 10$~nT and $25$~nT, respectively. 
From the model distributions in Figures \ref{modelBnfig}d--\ref{modelBnfig}g we calculate $210$~km and $150$~km, so equation (\ref{lp4}) is a reasonable approximation to the numerical model. As an example from observations, for shock 3 we estimate $l_{p,t} \approx 270$~km and $l_{p,r} \approx 260$~km assuming constant $V_{sh,n}$. Using the parameters in Table \ref{Tableshockprops} we calculate $l_{p,t} = 230$~km and $l_{p,r} = 210$~km using equations (\ref{lp1}) and (\ref{lp4}), in good agreement with observations. 
%For the alpha particles we estimate $l_{\alpha,t} \approx 510$~km, consistent with the expected $l_{p,t}/l_{\alpha,t} = 2$. 

When we consider a thermal distribution of protons, each proton will have different $v_c$ in equation (\ref{lp2}), with the spread in $v_c$ determined by upstream proton temperature and heating processes across the ramp. This results in a spread in $l_p$, which increases with $B_n$ according to equation (\ref{lp2}). The increased spread in $l_p$ result in more rapid gyrophase mixing, which smooths out the periodic fluctuations in the transmitted and reflected protons over a shorter distance downstream of the shock. By assuming a thermal speed $v_{th,d}$ just downstream of the ramp we spread in $l_p$, $\Delta l_p$, is obtained from equations (\ref{vnobl1}) and (\ref{lp2}) and given by
\begin{equation}
\Delta l_p = \frac{2 \pi B_n}{B_d^2 \Omega_{cp,d}} \left( B_n + B_{t1,d} \right) v_{th,d}.
\label{deltalp1}
\end{equation}
The number of oscillations observed before gyrophase mixing smooths them out can be approximated as $n = l_{p,t,r}/2 \Delta l_p$ \cite{balikhin2008}. For transmitted protons 
\begin{equation}
n \approx \frac{B_d^2 |V_{n,d}|}{2 B_n (B_n + B_{t1}) v_{th,d}},
\label{nlpeq}
\end{equation}
with $n \rightarrow \infty$ as $\theta_{Bn} \rightarrow 90^{\circ}$. 
For reflected protons $n$ will decrease due to $l_{p,r} < l_{p,t}$. The ratio of the distances required for reflected and transmitted protons to gyrophase mix is $l_{p,r}^2/l_{p,t}^2$, assuming the same $v_{th,d}$. For finite width shocks, the reflection point will vary with incoming proton speed, which will modify the speeds gained by reflection. This will likely increase $v_{th,d}$ for reflected protons compared with transmitted protons, further decreasing the distance required for gyrophase mixing of reflected protons at oblique shocks. The above analysis shows that for oblique shocks, the periodic fluctuations in reflected protons will relax more quickly than for transmitted protons resulting in distinct relaxation spatial scales, with the relaxation scales decreasing as $B_n$ increases or $\theta_{Bn}$ decreases. 

This analysis assumes laminar shocks with no wave activity. In practice, wave activity and fluctuations occur across the shocks, both in the electric and magnetic fields, is typical in observations, and are present in the five shocks studied here. These waves and fluctuations can contribute to ion heating and scattering, increasing the effective $v_{th,d}$ and thus decreasing $n$ compared with the laminar model predictions for oblique shocks. This could explain why gyrophase mixing in shock 3 develops more quickly than observed in the numerical model, Figures \ref{modelshockcomp}g--\ref{modelshockcomp}l. Similarly, in shocks 4 and 5 we do not observe any gyrophase bunching of transmitted protons, which are predicted based on the numerical model. Nevertheless, these results explain why the complex distributions associated with associated with shocks 1 and 2 persist far downstream, while an energetic halo population quickly forms for shock 3.

\subsection{Downstream Low-Frequency Magnetic Field Fluctuations}
\subsubsection{Observations}
\begin{figure*}[htbp!]
\begin{center}
\includegraphics[width=140mm, height=140mm]{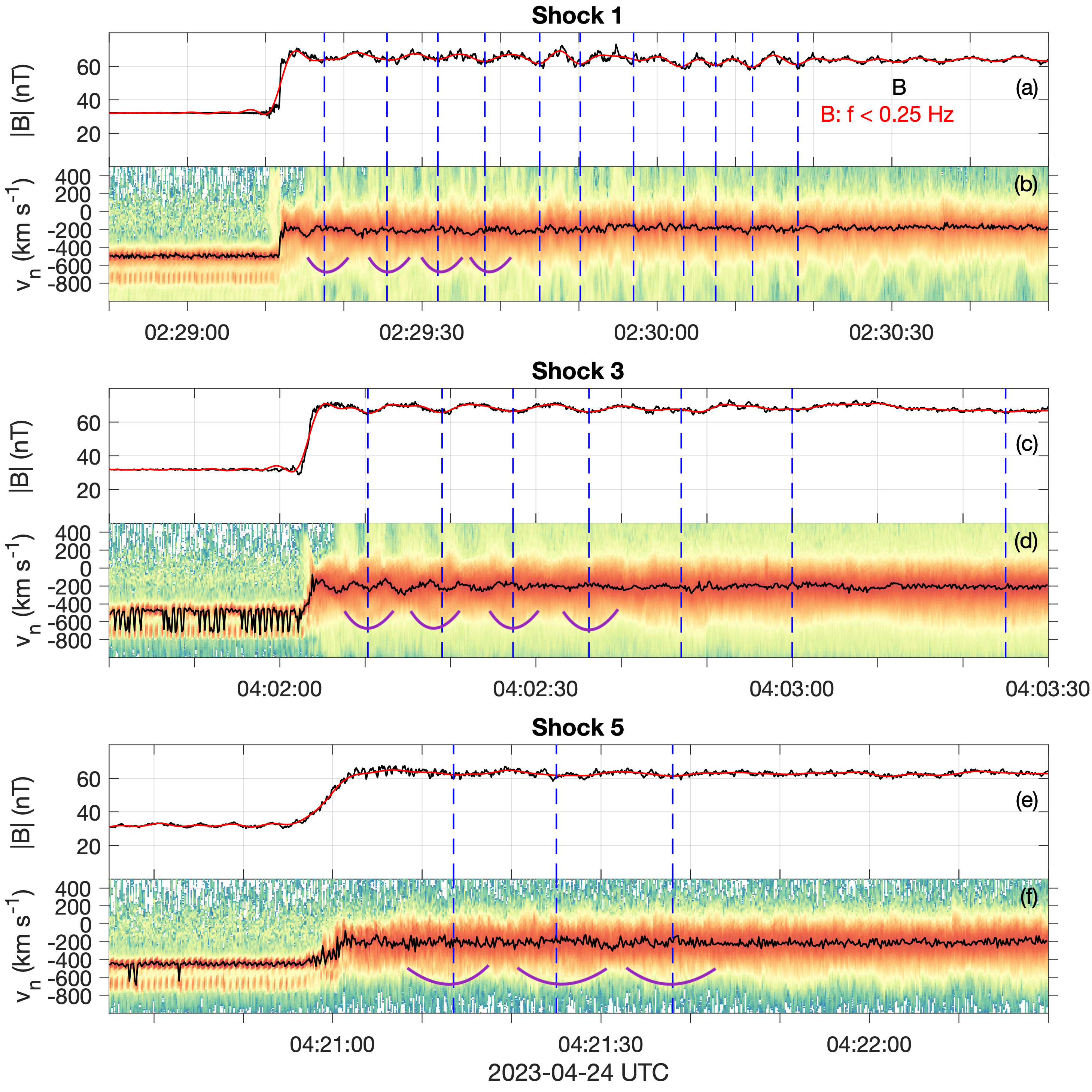}
\caption{Low-frequency compressional magnetic field fluctuations and the 1D reduced ion distributions along $v_n$ for shocks 1, 3, and 5. (a) The magnitude of the magnetic field $|{\bf B}|$ for shock 1. The black and red lines are $|{\bf B}|$ and $|{\bf B}|$ bandpass filtered below $0.25$~Hz, respectively. (b) 1D reduced ion distributions along $v_n$ of shock 1. (c)--(d) and (e)--(f) shocks 3 and 5 presented in the same format as shock 1. The purple lines indicate the approximate lower bound of $f_i(v_n)$ associated with He$^{2+}$. The blue vertical dashed lines correspond to times of minima in the low-frequency $|{\bf B}|$. }
\label{downstreamBfluct}
\end{center}
\end{figure*}

In this section we investigate the low-frequency compressional fluctuations observed downstream of the shocks, as seen in Figure \ref{overviewfig}. For all the shocks investigated here we observe quasi-periodic fluctuations in $|{\bf B}|$ downstream of the shocks. Figure \ref{downstreamBfluct} shows $|{\bf B}|$ and $f_i(v_n)$ for shocks 1, 3, and 5. In Figure \ref{downstreamBfluct} we have reduced the range in $v_n$ to focus on the transmitted proton and He$^{2+}$ populations. 

Figures \ref{downstreamBfluct}a and \ref{downstreamBfluct}b show $|{\bf B}|$ and $f_i(v_n)$ for shock 1. Figure \ref{downstreamBfluct}a shows that there are periodic fluctuations in $|{\bf B}|$ with frequency $f \approx 0.16$~Hz. This is more clearly seen by $|{\bf B}|$ bandpass-filtered below $0.25$~Hz. We also observe lower-amplitude higher-frequency compressional fluctuations. Figures \ref{downstreamBfluct}a and \ref{downstreamBfluct}b shows that at the minima in $|{\bf B}|$ there is an enhancement in $f_i(v_n)$ at $v_n = - 500$~km~s$^{-1}$. These enhancements correspond to the quasi-periodic motion of He$^{2+}$ downstream of the ramp. Upstream He$^{2+}$ are observed at $v_n \approx - 750$~km~s$^{-1}$ below the proton population, which is observed at $v_n \approx -500$~km~s$^{-1}$. Note that He$^{2+}$ speeds are overestimated [equation (\ref{vioneq})] Across the ramp and just downstream He$^{2+}$ are decelerated, such that the distributions overlap. In the downstream region He$^{2+}$ undergo quasi-periodic fluctuation along $v_n$, resulting in quasi-periodic fluctuations in $f_i(v_n)$, where He$^{2+}$ overlaps and are separated from the proton distribution \cite{graham2024}. This comparison shows that the minima in $|{\bf B}|$ correspond to approximately the largest $V_n$, or equivalently downstream $P_{dyn,\alpha}$, of He$^{2+}$. Similarly, the maxima in $|{\bf B}|$ correspond to the minima in downstream $P_{dyn,\alpha}$, where He$^{2+}$ overlaps with the protons. For shock 1, the largest amplitude quasi-periodic fluctuations in $|{\bf B}|$ are due the changes in $P_{dyn,\alpha}$. 

Figures \ref{downstreamBfluct}c and \ref{downstreamBfluct}d show the same data for shock 3. Like shock 1, we observe quasi-periodic fluctuations in $|{\bf B}|$, which are anti-correlated with increases in $P_{dyn,\alpha}$. Additionally, we observe fluctuations in $V_n$ associated with protons, with twice the frequency of the oscillations in the He$^{2+}$ population. However, it is difficult to resolve any clear $|{\bf B}|$ fluctuations with the same period as the protons. This confirms that the fluctuations in $|{\bf B}|$ are primarily due to He$^{2+}$. Toward the end of the interval the periods of $|{\bf B}|$ and He$^{2+}$ enhancements increases, which may result for $V_{sh}$ changing or gyrophase mixing of the He$^{2+}$ distributions. 

For shock 5, we similarly observe low-frequency $\delta |{\bf B}|$, and enhancements in He$^{2+}$ for $v_n < 0$ occurring where minima in $\delta |{\bf B}|$ are observed (Figures \ref{downstreamBfluct}e and \ref{downstreamBfluct}f). For shock 5 the enhancements in He$^{2+}$ are less pronounced than in shocks 1 and 3, and the period is significantly longer just behind the ramp. 

From the period of the most prominent fluctuations in $|{\bf B}|$ we estimate the wavelengths $l_B$ of these fluctuations using $V_{sh,n}$ for each shock to convert time to length. We also compare these values with the observed wavelengths of gyrophase bunched transmitted and reflected protons, and transmitted He$^{2+}$ and He$^+$, $l_{\alpha,t}$ and $l_{He,t}$, respectively. For transmitted protons we can only reliable estimate $l_{p,t}$ for shock 3. We also calculate the theoretical predictions of $l_{p,t}$ and $l_{p,r}$ using equations (\ref{lp1}) and (\ref{lp4}). The results are shown in Table \ref{Tablelengths}. The theoretical predictions, denoted by subscript $mod$ in Table \ref{Tablelengths}, are calculated for protons. For He$^{2+}$ and He$^+$ the predictions are twice and four times the proton values, respectively. 

For each of the shocks we find that $l_B$ is comparable to the observed wavelength $l_{\alpha,t}$ of He$^{2+}$, as expected from Figure \ref{downstreamBfluct}. For shocks 1 and 2 we estimate the wavelength $l_{He,t}$ of gyrophase bunched He$^+$. We find that $l_{He,t}$ is significantly larger than $l_B$, and we do not see any clear fluctuations at these scales. This suggests that He$^+$ densities are too low to produce any significant pressure fluctuations. Similarly, we find that for shocks 1--3 the observed and predicted $l_{p,r}$ are close to half the length of $l_B$. We note that the model and observed $l_{p,r}$ agree well. Only for shock 3 can we estimate $l_{p,t}$, which closely matches the predicted value. These results show that $l_B$ of the dominant fluctuations in $\delta {\bf B}$ are consistent with fluctuations in He$^{2+}$ pressure, and inconsistent with proton pressure fluctuations. 

\begin{table}
\begin{tabular}{ |p{2.5cm}||p{1.8cm}|p{1.8cm}|p{1.8cm}|p{1.8cm}|p{1.8cm}|  }
\hline
Parameter & Shock 1 & Shock 2 & Shock 3 & Shock 4 & Shock 5 \\
\hline
$l_B$ (km)              & 320 & 350 & 510 & 800 & 540 \\ 
$l_{p,t}$ (km)          & $-$ & $-$ & 270 & $-$ & $-$ \\ 
$l_{p,r}$ (km)          & 190 & 240 & 260 & $-$ & $-$ \\
$l_{p,t,mod}$ (km)      & 250 & 190 & 230 & 210 & 240 \\
$l_{p,r,mod}$ (km)      & 250 & 190 & 210 & 170 & 170 \\
$l_{\alpha,t}$ (km)     & 340 & 480 & 510 & 740 & 520 \\ 
$l_{He,t}$ (km)         & 700 & 1000 & $-$ & $-$ & $-$ \\ 
$n_{\alpha,u}/n_{p,u}$  & 0.08 & 0.09 & 0.05 & 0.09 & 0.04 \\ 
\hline
\end{tabular}
\caption{Estimated wavelengths of the low-frequency fluctuations downstream of the shock calculated magnetic field and wavelengths of the gyrobunched ions. Length scales are estimated by estimating the period of the fluctuations in time and converting to length using the $V_{sh,n}$ in Table \ref{Tableshockoverview}. Model predictions for the wavelengths associated with transmitted He$^{2+}$ and He$^+$ are twice and four times the magnitude of the proton prediction. }
\label{Tablelengths}
\end{table}

\subsubsection{Theoretical Explanation}
We now consider why the fluctuations in $|{\bf B}|$ develop. For low Mach number shocks the downstream region is relatively laminar so the total pressure downstream of the shock is expected to be relatively constant. Since the changes in $|{\bf B}|$ are correlated with the changes in the He$^{2+}$ distribution downstream of the shock, we investigate the relation between the changes in $|{\bf B}|$ and the He$^{2+}$ pressure fluctuations. 

The five shocks are characterized by low $\beta$, so the total downstream He$^{2+}$ pressure should be characterized by $P_{dyn,\alpha} \gg P_{\alpha}$ just downstream of the ramp. This is confirmed from the numerical model when He$^{2+}$ are used (not shown). The changes in $P_B$ should be balanced by the changes in $P_{dyn,\alpha}$ due to the gyromotion of He$^{2+}$ downstream of the shocks. By considering the peak-to-peak changes in $P_B$ and $P_{dyn,\alpha}$ downstream of the shocks we can estimate the He$^{2+}$ density required to generate the observed $\delta |{\bf B}|$. 

The magnetic field pressure is defined as $P_B = B^2/(2 \mu_0)$. In the limit where the perturbations in $|{\bf B}|$ are characterized by $\delta |{\bf B}| \ll |{\bf B}|$ the peak-to-peak change in $P_{B}$ is given by
\begin{equation}
\Delta P_{B,pp} \approx \frac{|{\bf B}| \, \delta |{\bf B}|_{pp}}{\mu_0},
\label{PBeq1}
\end{equation}
where $\delta |{\bf B}|_{pp}$ is the peak-to-peak change in $|{\bf B}|$. The He$^{2+}$ dynamic pressure is given by $P_{dyn,\alpha} = m_{\alpha} n_{\alpha}(n) V_{\alpha,n}(n)^2$, where $n_{\alpha}$ and $V_{\alpha,n}$ are the He$^{2+}$ density and speed along $\hat{\bf n}$. Downstream of the shock, $n_{\alpha}$ and $V_{\alpha,n}$ fluctuate with position along $\hat{\bf n}$ due to gyrophase bunching. We assume that the He$^{2+}$ flux is conserved along $\hat{\bf n}$ in the NIF. From flux conservation, it follows that the downstream He$^{2+}$ dynamic pressure is $P_{dyn,\alpha} = m_{\alpha} n_{u,\alpha} V_{n,u,\alpha} V_{n,d,\alpha}(n)$. Therefore, the peak-to-peak fluctuation in downstream He$^{2+}$ dynamic pressure is 
\begin{equation}
\Delta P_{dyn,\alpha,pp} \approx m_{\alpha} n_{u,\alpha} V_{n,u} \Delta V_{n,d,\alpha,pp},
\label{Pramaeq1}
\end{equation}
where we assume the upstream speed of He$^{2+}$ is equal to the protons. We assume that He$^{2+}$ are decelerated across the ramp by $\phi$, with bulk speed just downstream along $\hat{\bf n}$ given by equation (\ref{eqdeltavn}) using He$^{2+}$ mass and charge. Downstream of the shock, He$^{2+}$ gyrate around $V_{n,d}$, whence we obtain 
\begin{equation}
\Delta P_{dyn,\alpha,pp} \approx 2 m_{\alpha} n_{u,\alpha} V_{n,u} \left(  - \sqrt{V_{u,n}^2 - \frac{4 e \phi}{m_{\alpha}}} - V_{n,d} \right). 
\label{Pramaeq2}
\end{equation}
The estimated $\phi$ of these shocks is large, so $\phi$ plays an important role in determining the magnitude of the downstream He$^{2+}$ velocity and pressure fluctuations. We note that as $l$ increases the accuracy may decrease due to the use equation (\ref{eqdeltavn}), as illustrated by the changes in proton behavior as $l$ increases. 
By equating equations (\ref{PBeq1}) and (\ref{Pramaeq2}), we then obtain an estimate of the upstream He$^{2+}$ to proton number density ratio, given by 
\begin{equation}
\frac{n_{\alpha,u}}{n_{p,u}} \approx \frac{|{\bf B}| \, \delta |{\bf B}|_{pp}}{2 \mu_0 m_{\alpha} n_p V_{n,u} \left(  - \sqrt{V_{u,n}^2 - \frac{4 e \phi}{m_{\alpha}}} - V_{n,d} \right)},
\label{denrateq}
\end{equation}
where the upstream and downstream densities and speeds correspond to average values. From the measured $\delta |{\bf B}|_{pp}$, which we estimate from the peak and minimum in $|{\bf B}|$ just behind the ramp, and average upstream and downstream $V_n$ calculated in section \ref{shockoverviewsec}, we estimate $n_{\alpha,u}/n_{p,u}$, which are given in table \ref{Tablelengths}. From equation (\ref{denrateq}) we estimate $n_{\alpha,u}/n_{p,u}$ between $0.04$ and $0.09$, which are comparable to the nominal $n_{\alpha}/n_{p} \approx 0.04$ in the solar wind \cite{ogilvie1969,elliott2018}. The ratio $n_{\alpha}/n_{p}$ can be enhanced inside magnetic clouds, compared with typical solar wind conditions \cite{owens2018}. Over the time interval the five shocks are observed we obtain a median $n_{\alpha}/n_{p}$ of $\approx 0.08$ from OMNI data, consistent with our estimates. Based on the estimated $n_{\alpha}/n_{p}$, He$^{2+}$ contributes $\approx 14$~\% to $27$~\% of the total upstream dynamic pressure, which can have important implications for the energy partition across the shock. 

Assuming $n_{u,\alpha}/n_{u,p} = 0.08$ we can calculate the peak-to-peak fluctuations pressure fluctuations in the Liouville mapping model for the five shocks. We assume $T_{u,\alpha} = 12$~eV, although the total downstream pressure fluctuations are dominated by the dynamic pressure. We calculate the maximum peak-to-peak total pressures along $\hat{\bf n}$, $\Delta P_{nn,p}$ and $\Delta P_{nn,\alpha}$ downstream of the ramp. 
For shock 1, the predicted $\Delta P_{nn,p}$ and $\Delta P_{nn,\alpha}$ are comparable. For shocks 2--4, $\Delta P_{nn,\alpha}$ ranges from $\approx 15$~\% (shock 2) to $\approx 60$~\% (shock 3) of $\Delta P_{nn,p}$. This would predict that the dominant $\delta |{\bf B}|$ should be at the proton gyrophase bunching scale, rather than the He$^{2+}$ scale. A likely explanation for the discrepancy is that the model is not capturing some physical processes affecting protons across the ramp and in the downstream region. For example, proton scattering by waves and turbulent $\delta {\bf B}$ could suppress proton gyrophase bunching downstream of the transmitted protons, substantially reducing $\Delta P_{nn,p}$. This was illustrated for shock 5 in Figure \ref{modelshockcomp}, where gyrophase bunching of protons is clear in the model, but is not observed by MMS. 

We conclude that the most prominent quasi-periodic $\delta |{\bf B}|$ are generated by fluctuations in $P_{dyn,\alpha}$. These results suggest that downstream $\delta |{\bf B}|$ due to fluctuations in $P_{dyn,\alpha}$ should often be observed at low Mach number low $\beta$ planetary bow shocks and interplanetary shocks. 

\section{Conclusions} \label{conclusionsec}
We have investigated the properties of the five quasi-perpendicular bow shock crossings observed on 24 April 2023. The shocks were observed as a magnetic cloud behind a CME crossed Earth, resulting in low Mach number, low beta bow shock crossings. The bow shock crossings exhibited similar Mach numbers, but had varying shock widths and shock-normal angles. The observation of these shocks by MMS provide a unique opportunity to study variations in shock structure and how this modifies the kinetic behavior of ions across these shocks.

The key results of this paper are: 
\begin{enumerate}
    \item The shock width $l$ increases as the shock-normal angle $\theta_{Bn}$ decreases for comparable Mach numbers. 
    \item For a given set of upstream parameters, the fraction of reflected protons decreases as $l$ increases. For the shocks studied here, proton reflection is suppressed when $l$ becomes comparable to $d_p$ or $V_n/\Omega_{cp}$. This suggests at low Mach number shocks proton reflection is favored for nearly perpendicular shocks due to the relation between $l$ and $\theta_{Bn}$ and accounts for observed proton reflection in shocks 1--3 and lack of proton reflection in shocks 4 and 5. 
    \item For nearly perpendicular shocks the reflected protons remain phase bunched far downstream of the shock, meaning the reflected protons are found in localized regions of velocity space. This results in quasi-periodic striations in the 1D reduced distribution functions.
    \item As $\theta_{Bn}$ decreases reflected protons undergo gyrophase mixing, which results in the distributions of reflected protons forming ring-like distributions. The rate of gyrophase mixing increases as $\theta_{Bn}$ decreases.  
    \item Downstream periodic fluctuations in the magnitude of the magnetic field result from periodic fluctuations in the downstream dynamic pressure of He$^{2+}$. 
\end{enumerate}

These results have important implications for low Mach number planetary bow shocks and interplanetary shocks, for example, shocks associated with co-rotation interaction regions are characterized by low Mach numbers. In these regions the complex ion dynamics typically cannot be resolved due to the limited temporal resolution of particle instruments and the fact that interplanetary shocks move past the spacecraft much faster than planetary bow shocks. These results show that the shock width and shock-normal angle of low Mach number shocks are crucial in determining the ion dynamics and overall ion properties, such as temperature and pressure, downstream of the shock. 

\section*{Data Availability Statement}
MMS data are available at \url{https://lasp.colorado.edu/mms/sdc/public} and \url{https://spdf.gsfc.nasa.gov/pub/data/mms/}. We use survey and burst mode magnetic field data from FGM \cite{mms1fgmbrst,mms1fgmsrvy}, electric field data from EDP \cite{mms1edpdcebrst}. For ion data we use fast and burst mode distributions from FPI \cite{mms1fpidisdistfast,mms1fpidisdistbrst}, burst mode moments from FPI \cite{mms1fpidismomsbrst}. For electron data we use fast and burst mode electron distributions from FPI \cite{mms1fpidesdistfast,mms1fpidesdistbrst} and electron moments from FPI \cite{mms1fpidesmomsbrst}. The data analysis was performed using the irfu-matlab software package \cite{khotyaintsev_2024_11550091}.
The scripts required to generate the model data and the figures in this paper are available at \url{https://doi.org/10.5281/zenodo.13732515} \cite{graham_2024b}. 

\acknowledgments
We thank the entire MMS team and instrument PIs for data access and support. This work was supported by the Swedish National Space Agency (SNSA), Grant 206/19. 

\appendix
\section{Numerical Model} \label{app1}
To model the shocks we assume profiles of $B_{t1}$, $n_e$, and $P_e$ based on observations, which are given by \cite{graham2024}
\begin{equation}
B_{t1}(n) = - B_0 \tanh{\left(\frac{n}{l} \right)} + B_1,
\label{Bt1model}
\end{equation}
\begin{equation}
n_{e} = - n_0 \tanh{\left(\frac{n}{l} \right)} + n_1,
\label{nemodel}
\end{equation}
\begin{equation}
P_{e}(n) = - P_0 \tanh{\left(\frac{n}{l} \right)} + P_1,
\label{Pemodel}
\end{equation}
where $l$ is the characteristic ramp width, and $B_0$, $B_1$, $n_0$, $n_1$, $P_0$, and $P_1$ determine the upstream and downstream conditions. We assume $B_n$ is constant and $B_{t2} = 0$~nT. The upstream and downstream values are, for example, given by $B_{t1,u} = B_1 - B_0$ and $B_{t1,d} = B_1 + B_0$. Equations (\ref{Bt1model})--(\ref{Pemodel}) provide a reasonable approximation to the observed shock profiles.

The current density ${\bf J}$ is determined from $B_{t1}(n)$ using Ampare's law: 
\begin{equation}
J_{t2}(n) = - \frac{B_0 \mathrm{sech}^2 \left( \frac{n}{l} \right)}{\mu_0 l},
\label{Jt2model}
\end{equation}

The three components of the electric field are calculated from the generalized Ohm's law:
\begin{equation}
{\bf E} = - {\bf V}_i \times {\bf B} + \frac{{\bf J} \times {\bf B}}{e n_e} - \frac{\nabla \cdot {\bf P}_e}{e n_e},
\label{GOLmodel}
\end{equation}
where ${\bf P}_e$ is the electron pressure tensor. To simplify the calculation we assume a scalar electron pressure $P_e$ when determining the contribution to ${\bf E}$. 

The three components of the electric field are 
\begin{equation}
E_n(n) = - \frac{1}{e n_e(n)} \left( J_{t2}(n) B_{t1}(n) + \frac{\partial P_e(n)}{\partial n} \right),
\label{Enmodel}
\end{equation}
\begin{equation}
E_{t1}(n) = \frac{J_{t2}(n) B_n}{e n_e(n)},
\label{Et1model}
\end{equation}
\begin{equation}
E_{t2} = - V_{u,n} B_{t1,u},
\label{Et2model}
\end{equation}
in the normal-incidence frame (NIF) of the shock, defined as the frame in which the shock is stationary and upstream flow is normal to the shock. In equation (\ref{Et2model}) we assume $E_{t2}$ is constant across the shock. The cross-shock potential in the NIF is obtain by integrating $E_n$ over $n$ and is given by
\begin{equation} \label{phieq}
\begin{split}
\phi & = \frac{2 B_0^2}{e \mu_0 n_0} + \frac{B_0 (B_1 n_0 - B_0 n_1)}{e \mu_0 n_0^2} \log{\left( \frac{n_1 + n_0}{n_1 - n_0} \right)} + \frac{P_0}{e n_0} \log{\left( \frac{n_1 + n_0}{n_1 - n_0} \right)}\\
 & = \frac{(B_{t1,d} - B_{t1,u})^2}{e \mu_0 (n_d - n_u)} + 
 \frac{(B_{t1,d} - B_{t1,u})(B_{t1,u} n_d - B_{t1,d} n_u)}{e \mu_0 (n_d - n_u)^2} \log{\left( \frac{n_d}{n_u} \right)} + \frac{(P_{e,d} - P_{e,u})}{e (n_d - n_u)} \log{\left( \frac{n_d}{n_u} \right)}.
\end{split}
\end{equation}
Here $\phi$ is expressed in terms of the quantities in equations (\ref{Bt1model})--(\ref{Pemodel}) and the upstream and downstream parameters. In this model $\phi$ is independent of $l$. 

To model the ion dynamics across the shocks we use Liouville mapping, which states that the particle's probability density is conserved along the particle trajectories, and can be expressed as
\begin{equation}
f_1({\bf x}_1,{\bf v}_1,t_1) = f_0({\bf x}_0,{\bf v}_0,t_0), 
\label{lmappingeq}
\end{equation}
where the subscripts $0$ and $1$ refer to the modelled and upstream distribution. We assume the upstream distribution to be Maxwellian and given by
\begin{equation}
f_i({\bf v}) = \frac{n_i}{\pi^{3/2} v_i^3} \exp{\left( - \frac{(v_n - V_{u,n})^2 + v_{t1}^2 + v_{t2}^2}{v_i^2} \right)},
\label{Distmodel}
\end{equation}
where $v_i = \sqrt{2 k_B T_i/m_i}$ is the ion thermal speed and $m_i$ is the ion mass. We assume the solar wind is uniform in position and time upstream of the shock and that ${\bf E}$ and ${\bf B}$ do not change with time. In this case equation (\ref{lmappingeq}) is independent of $t$ and $f_1({\bf v})$ depends only on the $n$ position. 

To approximate equation (\ref{lmappingeq}) and calculate $f_1(n,{\bf v}_1)$ we define a grid in velocity space with spacing of $10$~km~s$^{-1}$, along $v_n$, $v_{t1}$, and $v_{t2}$. Similarly, we define a spatial grid along $n$ with separations of $10$~km. To calculate $f_i({\bf v})$ on this grid of ${\bf v}$ and $n$ we use a large number of test particles to construct the distribution functions. The test particles are initialized in the solar wind at $n = 2000$~km and the particle velocity and position is calculated using the equations of motion 
\begin{equation}
\frac{d {\bf x}}{d t} = {\bf v}, \frac{d {\bf v}}{d t} = \frac{Z e}{m_i} \left( {\bf E} + {\bf v} \times {\bf B} \right),
\label{eqsofmotion}
\end{equation}
where $Z$ is the charge number of the ions, and ${\bf E}$ and ${\bf B}$ are given by equations (\ref{Bt1model}) and (\ref{Enmodel})--(\ref{Et2model}). The test particles are distributed such that the initial speeds corresponds to a Maxwellian distribution given by equation (\ref{Distmodel}). The contribution of each test particle to $f_{i}(n,{\bf v})$ is calculated as the dwell time of the particle in each point in the grid. To simplify calculating the dwell time, the particle positions and velocities are interpolated to uniform time steps of $\delta t = 10^{-3}$~s. The dwell times are then simply histograms of counts in the elements of the velocity-space grid. The histograms are calculated in 1D and 2D velocity space to produce 1D and 2D reduced distribution functions. As $v_n$ decreases the number of counts increases at a given normal position, which enforces the conservation of the particle flux $n_i V_n$ along $\hat{\bf n}$ throughout the domain. 

Additionally, each of the particles is assigned a weight $W$ when constructed the histograms, which is given by
\begin{equation}
W = \frac{v_h^3}{v_i^3} \exp{\left( \frac{v^2 (v_i^2 - v_h^2)}{v_i^2 v_h^2} \right)},
\label{weightsmodel}
\end{equation}
where $v$ is the initial speed of the particle in the model solar wind frame, $v_h$ is the thermal speed of the initialized distribution of test particles and $v_i$ is the thermal speed of the distribution we are modelling [equation (\ref{Distmodel})]. Here, we use $v_h > v_i$ to increase the proportion of test particles at suprathermal speeds to more accurately model the small fraction of protons reflected at the shock ramp. To convert the histograms of weighted particle counts to physical 1D and 2D reduced distributions we normalize the 1D and 2D histograms using $f_{1D}(n,{\bf v}) = n_{p,0} H_{1D}(n,{\bf v})/\int H_{1D,0}({\bf v}) dv$ and $f_{2D}(n,{\bf v}) = n_{p,0} H_{2D}(n,{\bf v})/\int H_{2D,0}({\bf v}) d^2v$, where the integral is over the 1D and 2D histograms at $n = 500$~km and $n_{p,0}$ is the density of the upstream ion distribution used in equation (\ref{Distmodel}).

We primarily focus on the proton distributions. We assume the upstream proton temperature is $T_p = 3$~eV and use $T_h = 12$~eV to initialize the distribution of test particles and calculate $W$. To calculate the distributions we use $10^5$ test particles. For each shock we calculate the proton distributions in the domain $-2000$~km $\leq n \leq 500$~km, where $n = 0$ is the center of the shock ramp.

%\bibliography{magrecpapers}

\end{document}